\documentclass[fleqn,usenatbib]{mnras} 

\usepackage{newtxtext,newtxmath}

\usepackage[T1]{fontenc}

\usepackage[normalem]{ulem}

\DeclareRobustCommand{\VAN}[3]{#2}
\let\VANthebibliography\thebibliography
\def\thebibliography{\DeclareRobustCommand{\VAN}[3]{##3}\VANthebibliography}

\usepackage{graphicx}	
\usepackage{amsmath}	
\usepackage{enumitem}
\usepackage{float}
\usepackage{xcolor}

\newcommand{\kms}{km s$^{-1}$}
\newcommand{\coude}{Coud{\'e}}

\hypersetup{draft}

\title{A 2+1+1 quadruple star system containing the most eccentric, low-mass, short-period, eclipsing binary known}

\author[E. Han et al.]{
E.~Han,$^{1}$\thanks{E-mail: eunkyu.han@utexas.edu (EH)}
S.A.~Rappaport,$^{2}$
A.~Vanderburg,$^{3}$
B.M.~Tofflemire,$^{1}$\thanks{51 Pegasi b fellow}
T.~Borkovits,$^{4,5,6}$
H.\,M.\,Schwengeler,$^{7}$
\newauthor
P.~Zasche,$^{8}$
D.M.~Krolikowski,$^{1}$
P.S.~Muirhead,$^{9}$
M.\,H.\,Kristiansen,$^{10,11}$
I.A.~Terentev,$^{7}$
M.\,Omohundro,$^{7}$
\newauthor
R.~Gagliano,$^{12}$
T.\,Jacobs,$^{13}$
D.\,LaCourse$^{14}$
\\
$^{1}$Department of Astronomy, The University of Texas at Austin, 2515 Speedway, Stop C1400, Austin, TX 78712, USA\\
$^{2}$Department of Physics, and Kavli Institute for Astrophysics and Space Research, M.I.T., Cambridge, MA 02139, USA\\
$^{3}$Department of Astronomy, The University of Wisconsin-Madison, 475 N.\,Charter St., Madison, WI 53706, USA \\
$^{4}$Baja Astronomical Observatory of University of Szeged, H-6500 Baja, Szegedi \'ut, Kt. 766, Hungary\\
$^{5}$Konkoly Observatory, Research Centre for Astronomy and Earth Sciences,  H-1121 Budapest, Konkoly Thege Mikl\'os \'ut 15-17, Hungary\\
$^{6}$ELTE Gothard Astrophysical Observatory, H-9700 Szombathely, Szent Imre h. u. 112, Hungary \\
$^{7}$Citizen Scientist, Planet Hunter, Petrozavodsk, Russia \\
$^{8}$ Astronomical Institute, Charles University, Faculty of Mathematics and Physics, V Hole\v{s}ovi\v{c}k\'{a}ch 2, CZ-180 00, Praha 8, Czech Republic \\
$^{9}$Department of Astronomy \& Institute for Astrophysical Research, Boston University, 725 Commonwealth Avenue, Boston, MA 02215, USA\\
$^{10}$DTU Space, National Space Institute, Technical University of Denmark, Elektrovej 327, DK-2800 Lyngby, Denmark \\
$^{11}$Brorfelde Observatory, Observator Gyldenkernes Vej 7, DK-4340 T\o ll\o se, Denmark \\
$^{12}$Amateur Astronomer, Glendale, AZ 85308 \\
$^{13}$Amateur Astronomer, 7507 52nd Place NE Marysville, WA 98270, USA \\
$^{14}$Amateur Astronomer, 12812 SE 69th Place Bellevue, WA 98006, USA \\
}

\date{Accepted 2021 November 29}

\pubyear{2021}

\begin{document}
\label{firstpage}
\pagerange{\pageref{firstpage}--\pageref{lastpage}}
\maketitle

\begin{abstract}
We present an analysis of a newly discovered 2+1+1 quadruple system with {\em TESS} containing an unresolved eclipsing binary (EB) as part of TIC 121088960 and a close neighbor TIC 121088959. The EB consists of two very low-mass M dwarfs in a highly-eccentric ($e$ = 0.709) short-period ($P$ = 3.04358 d) orbit. Given the large pixel size of {\em TESS} and the small separation ($3\farcs9$) between TIC 121088959 and TIC 121088960 we used light centroid analysis of the difference image between in-eclipse and out-of-eclipse data to show that the EB likely resides in TIC 121088960, but contributes only $\sim$10\% of its light.  Radial velocity data were acquired with iSHELL at NASA's Infrared Facility and the \coude \,spectrograph at the McDonald 2.7-m telescope. For both images, the measured RVs showed no variation over the 11-day observational baseline, and the RV difference between the two images was $8 \pm 0.3$ km s$^{-1}$. The similar distances and proper motions of the two images indicate that TIC 121088959 and TIC 121088960 are a gravitationally bound pair.  Gaia's large RUWE and astrometric\_excess\_noise parameters for TIC 121088960, further indicate that this image is the likely host of the unresolved EB and is itself a triple star. We carried out an SED analysis and calculated stellar masses for the four stars, all of which are in the M dwarf regime: 0.19 M$_\odot$ and 0.14 M$_\odot$ for the EB stars and 0.43 M$_\odot$ and 0.39 M$_\odot$ for the brighter visible stars, respectively. Lastly, numerical simulations show that the orbital period of the inner triple is likely the range 1 to 50 years.
  \end{abstract}




\section{Introduction}

\indent Stellar companions are a common product of star formation and hence studying stellar multiplicity and the associated properties (e.g. period and mass ratio distributions) can provide pivotal insights into understanding the nature of star formation processes and stellar evolution. Since the first systematic studies of the observational properties of close triple systems \citep[][]{Fekel1981}, efforts have been made to search for and compile catalogs of multiple systems \citep[e.g. see][ and references therein]{Tokovinin1997, Tokovinin2008, Tokovinin2014a, Eggleton2009, Raghavan2010, Rappaport2013, Borkovits2016} as well as to conduct statistical studies of them \citep[e.g.][]{Duchene2013, Winters2019}. These studies found that multiple star systems are common in our galaxy. Multiplicity of main-sequence solar-type stars (M$_*$ $\approx$ $0.7-1.3$ M$_{\odot}$) is $41 \pm 3\%$ \citep[][]{Raghavan2010} and that of low-mass stars (M$_*$ $\approx$ $0.1-0.6$ M$_{\odot}$) is $26 \pm 3\%$ \citep[][]{Winters2019}. Although binaries are the most common type among multiples, triple and higher-order systems take up considerable fractions; $\sim$25\% of solar-type multiples \citep[][]{Eggleton2008} and $\sim$21\% of low-mass multiples \citep[][]{Reid2000} have 3 or more components.

The majority of observed multiples are in hierarchical systems. For triple systems, there is only one dynamically stable configuration, where an inner binary is orbited by a third body (2+1 configuration) with a ratio between the triple and binary periods typically exceeding 5-10 \citep[see, e.~g.][]{Mardling2001}. For quadruple star systems, there are two possible configurations. One involves two binary systems orbiting each other's center of mass (2+2 configuration), while the other involves a hierarchical triple system orbited by a fourth companion (2+1+1 configuration). \citet[][]{Tokovinin2014b} notes that the 2+2 systems and the 2+1+1 systems may well form via different mechanisms. Because these two configurations likely have a different history, and may even involve a different star formation mechanism, searching for and characterizing both types of quadruple systems are likely to be rather important to our overall understanding of the astrophysics of star formation.

The long-term dynamical evolution of hierarchical multiples after they have already formed is also complex process. A prominent mechanism involved in the long-term evolution are the von Zeipel-Lidov-Kozai (ZLK) oscillatory cycles \citep[][]{vonZeipel1910, Lidov1962, Kozai1962}, which transfer angular momentum between the inner and the outer orbits. This can result in cycles of enhanced eccentricity and orbital plane tilts of the inner binary. Multiple groups have extensively studied the ZLK cycles and their effect on orbital evolution of multiple systems. Theoretical studies such as \citet[][]{Kiseleva1998}, \citet[][]{Fabrycky2007}, and \citet[][]{Naoz2014} have shown that ZLK cycles with tidal friction (ZCTF), can shrink the orbits of the inner binaries in triple systems. Several studies have searched for and shown observational evidence for such hierarchical triples with close outer orbits \citep[e.g.][]{Rappaport2013, Borkovits2016, Hajdu2017, Borkovits2020}. In addition, studies like \citet[][]{Pejcha2013}, \citet[][]{Hamers2015}, \citet[][]{Vokrouhlicky2016}, and \citet[][]{Hamers2017} explored the ZLK oscillations using N-body simulations and found that quadruples are more likely to have high inner orbital eccentricities than triples. To our knowledge, this latter prediction has not yet been confirmed by observations.

Despite the efforts and the advancement of both theoretical and observational techniques, the dominant formation mechanism of short-period eccentric binaries is still not clearly known. A recent study showed that the KCTF itself cannot explain the large number of close binary systems and, especially, the frequency of pre-MS close binary stars \citep[][]{Moe2018}. Previous studies such as those of \citet[][]{Tokovinin2008} showed an enhancement of the inner period distribution at a few days among both triple and quadruple systems, which was thought to be a product of ZLK oscillations. However, more recently \citet{Tokovinin2020} has shown that this was merely the consequence of an observational selection effect, since at the time most multiple system were discovered amongst eclipsing binaries, with a naturally strong bias toward short periods. Nowadays, however, the majority of inner subsystems are discovered spectroscopically and, therefore, according to the up-to-date edition of the Multiple Star Catalog \citep[MSC,][]{Tokovinin2018}, the cumulative inner-period distribution of multiple systems is smooth.

Several observational and statistical efforts have been made to study occurrence rates of the 2+2 and the 2+1+1 quadruple systems. \citet[][]{Raghavan2010} searched through a distance-limited sample of 454 solar-type stars within 25\,pc and found 11 quadruples, among which 9 are 2+2 systems and only 2 are 2+1+1 systems. Using a volume-limited sample of 4847 solar-type stars in 67 pc from \citet[][]{Tokovinin2014a}, \citet[][]{Tokovinin2014b} carried out a statistical study and calculated that 74\% of quadruples are 2+2 systems. Similar findings are seen in the most recent edition of the MSC that among nearly 500 quadruple systems discovered-to-date, 23\% of the systems are 2+1+1 and the other 77\% are 2+2 systems. All studies point out that the 2+1+1 systems are less frequent than the 2+2 systems.

In this paper, we report the discovery of a hierarchical 2+1+1 quadruple stellar system discovered with NASA's Transiting Exoplanet Survey Satellite ({\em TESS}) Mission. All four stars are low-mass main-sequence stars with masses in the range of 0.14--0.43 M$_\odot$. The inner triple consists of a highly-eccentric short-period EB as an unresolved component of TIC 121088960. The outer quadruple is formed by TIC 121088959 with TIC 121088960 at a projected separation of $\sim3\farcs9$ at a distance of $\sim83$\,pc, giving a projected physical separation of $\sim$320 AU.

This paper is organized as follows. In Section 2, we describe the details of the data we used in the analysis. In Section 3, we present our analysis, modeling procedure, and results. In Section 4, we establish our reasoning for the host of EB. In Section 5, we discuss in detail the motivation for, and architecture of, a 2+1+1 system as well as the parameters of the constituent stars and their orbits. In Section 6, we present our analysis on the parameters of the inner triple. In section 7, we compare our highly-eccentric EB with other low-mass eccentric EBs. In Section 8, we investigate the effect of ZLK oscillations on the orbits of the constituent stars. Finally, in Section 9, we summarize our results and draw some final conclusions.

\section{Discovery, Data, and Observations}

\subsection{Discovery}
\label{sec:discovery}

Work by a group of amateur astronomy `surveyors', who appear on this paper and call themselves the `visual survey group' (VSG), have made a number of unexpected discoveries using the {\em Kepler} \citep{Borucki2010}, {\em K2} \citep{Howell2014}, and {\em TESS} data sets \citep{Ricker2015}.  These VSG discoveries were summarized by \citet{Rappaport2019}, and have continued with the discovery of a new class of `tidally tilted' pulsators\footnote{These are systems where one of the star's pulsation axis has been tilted into the orbital plane of the binary.  This then allows the observer to view the star from a complete range of latitudes.} in eclipsing binaries (\citealt{Handler2020}; \citealt{Kurtz2020}).

Upon the data release of each {\em TESS} sector's Candidate Target List \citep[CTL,][]{Stassun2018}, the Pre-search Data Conditioned Simple Aperture Photometry (PDCSAP) lightcurves, binned at 30-minute cadence, are displayed with {\sc LcTools} \citep{Schmitt2019}. The {\em TESS} data are hosted by the Mikulski Archive for Space Telescopes (MAST) and are downloaded as FITS files.  Among other features, {\sc LcTools} supports a variety of data retrieval options and it provides an efficient search method for visual lightcurve surveys.

VSG identified TIC 121088959/60 as containing a highly eccentric EB in Sector 31 (S31) with a ratio of eclipse spacing of 1:11 corresponding to an eccentricity of $e \gtrsim 0.7$ (discussed in Sects. \ref{sec:tess} and \ref{sec:analysis}). After TIC 121088959/60 was identified, we generated a 2-minute short-cadence CTL lightcurve. We also found an additional Full Frame Image (FFI) data set in Sector 4 (S4) using the Web TESS Viewing Tool (WTV); this extends the observation baseline to 2 years.

Subsequently, we inspected a $15 \times 15$ pixel FFI cutout centered on TIC 121088959/60 in S31 with Lightkurve \citep{LKC}. We found the eclipses to be on-target, but, in fact, we have no way of knowing initially which stellar image hosts the eclipses from the {\em TESS} data alone. Finally, we used a custom pipeline to reduce the S4 FFI data.

\subsection{{\em TESS} Observations}
\label{sec:tess}
\begin{figure}
\includegraphics[width=\columnwidth]{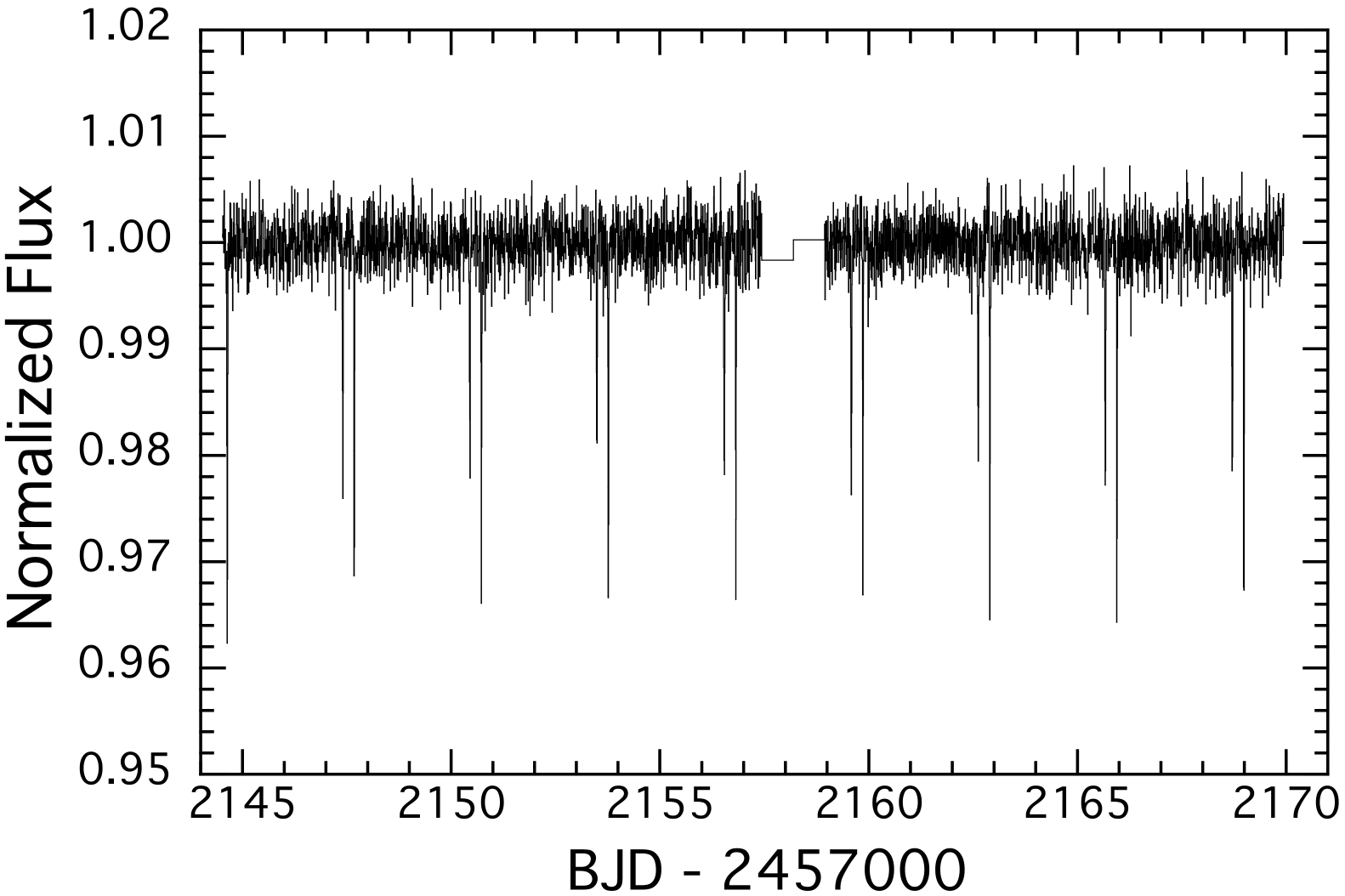}
\vspace{-5mm}
\caption{The {\em TESS} Sector 31 lightcurve for the combined light of TIC 121088959 and TIC 121088960.  The set of highly eccentric eclipses with a $\sim$3.04-day period is apparent.}
\label{fig:fullLC}
\end{figure}

\indent The S31 lightcurve in 10-minute bins is shown in Figure~\ref{fig:fullLC}. Immediately evident is the set of highly eccentric eclipses with a 3.04-day period.  Using data from both S4 and S31 we derive a period of 3.04358 days.

\begin{figure}
\includegraphics[width=1\columnwidth]{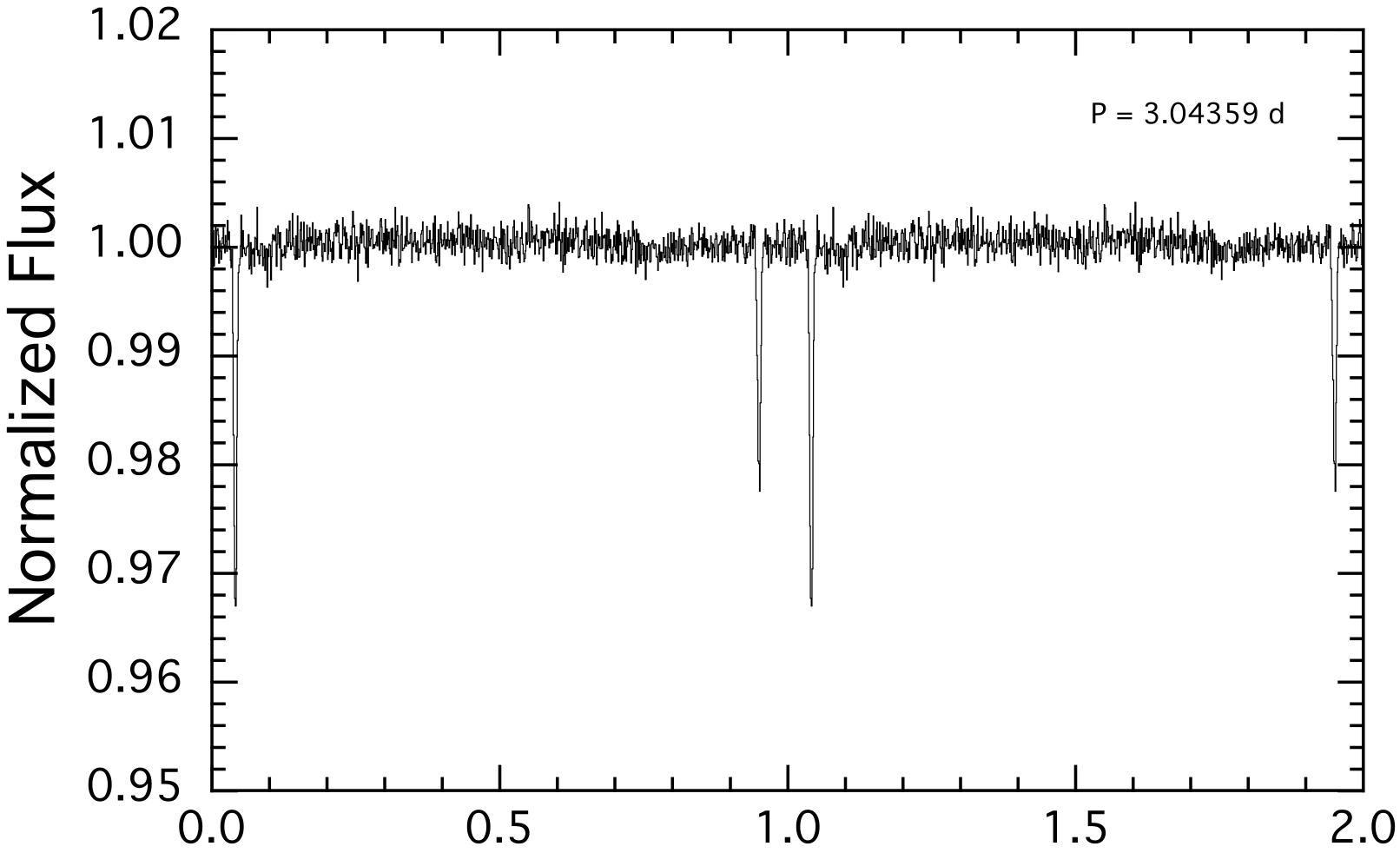}
\includegraphics[width=1\columnwidth]{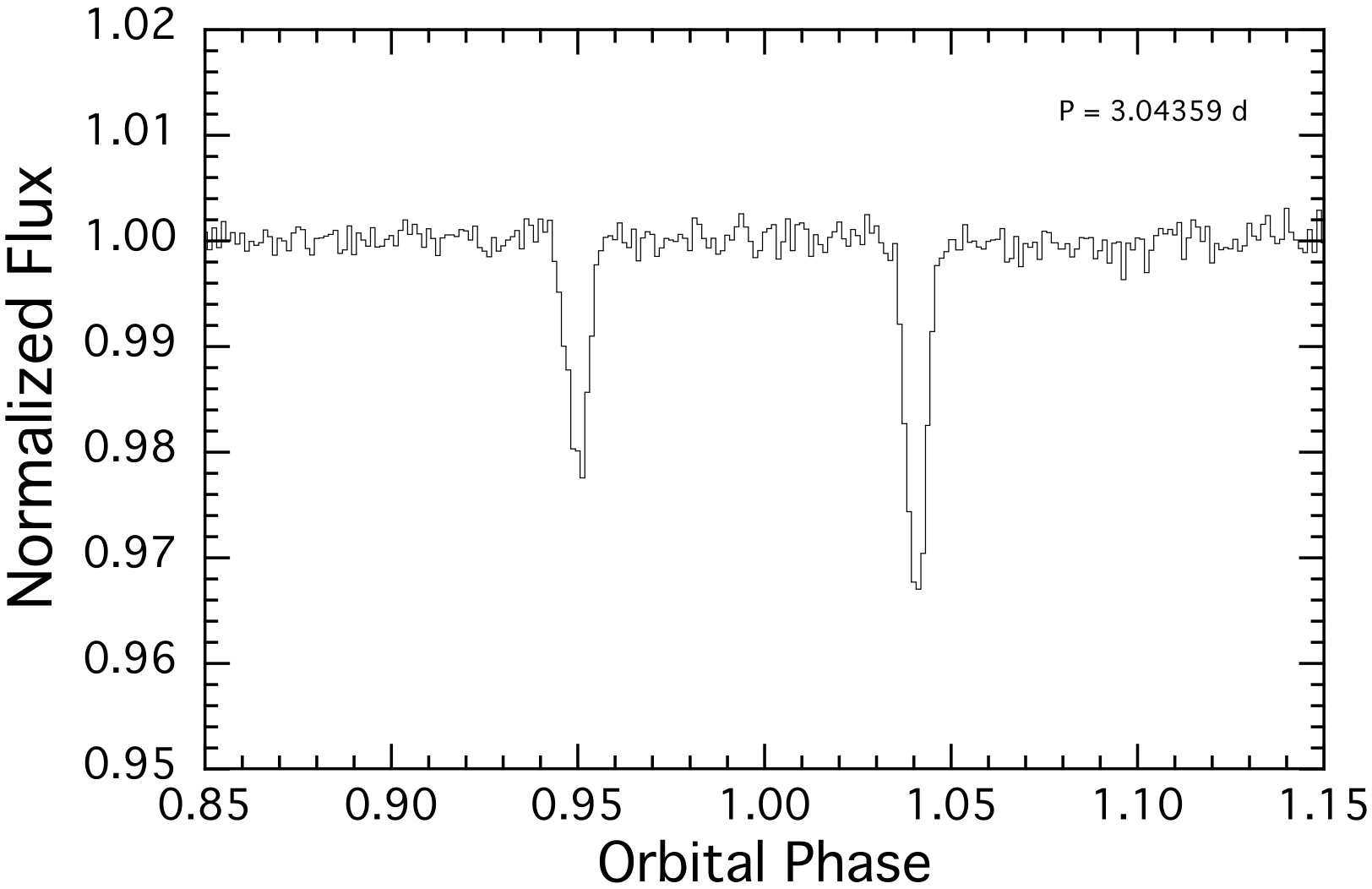}
\vspace{-5mm}
\caption{Phase folded lightcurve of the {\em TESS} Sector 31 data for the combined light of TIC 121088959 and TIC 121088960 about the 3.04358-day orbital period. Phase zero is defined here as the approximate time of periastron passage. Each bin is 4 minutes in duration. The bottom panel, which is a zoom-in on the top panel, shows that both eclipse have close to the same duration indicating that the argument of periastron is close to either 0$^\circ$ or 180$^\circ$.}
\label{fig:fold}
\end{figure}

\indent We then produced a phase folded lightcurve which is shown in Figure~\ref{fig:fold}. A visual inspection of the lightcurve shows (i) no discernible out of eclipse modulations in flux such as might be due to ellipsoidal light variations, and (ii) the primary and secondary eclipses to have close to the same duration.

\begin{figure}
\includegraphics[width=0.9\columnwidth]{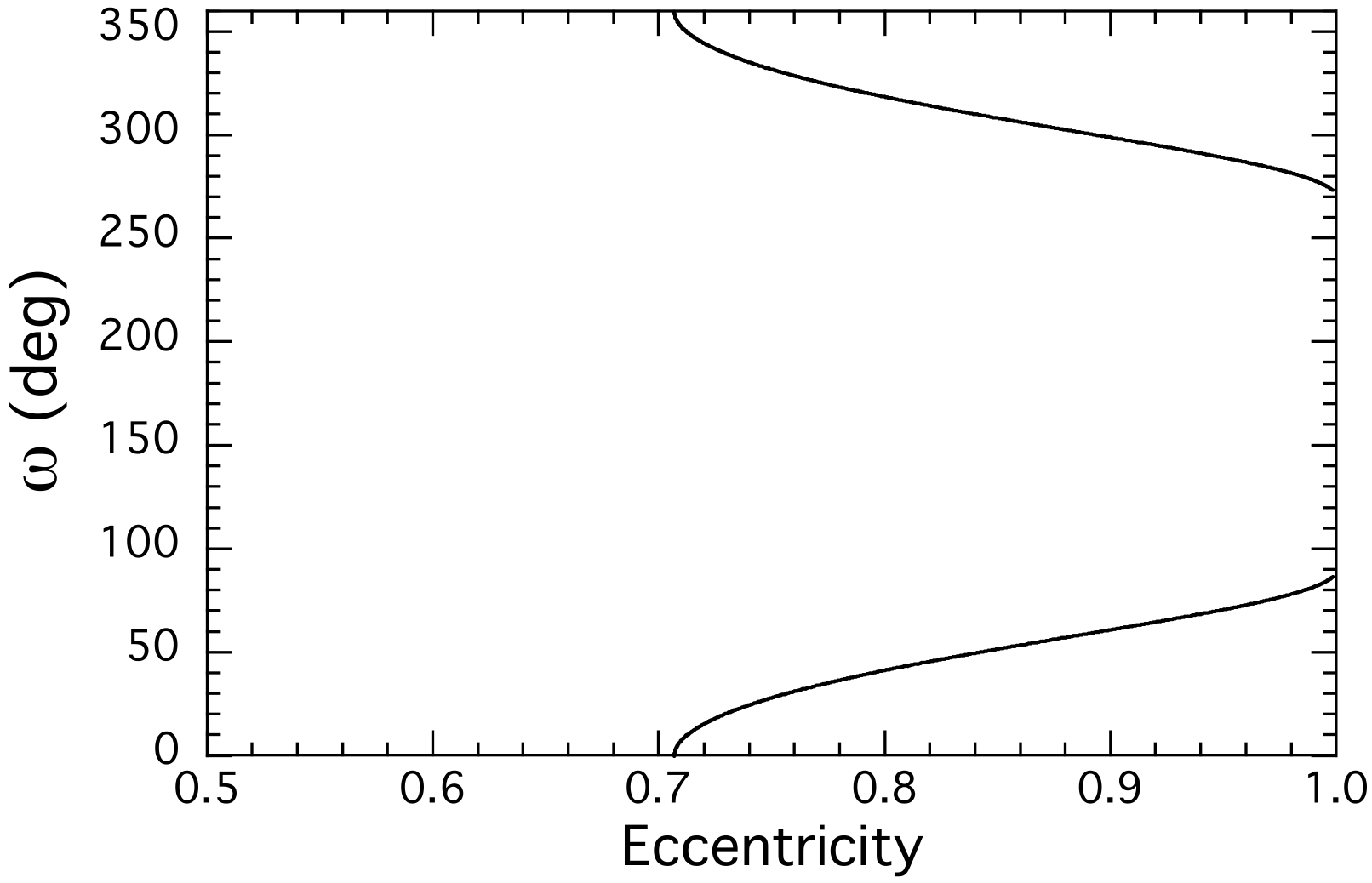}
\caption{Solutions of Eqn.~(\ref{eqn:sterne}) showing allowed values of $e$ for each possible value of $\omega$ for the observed difference in phase between the two eclipses.}
\label{fig:e_P_contour}
\end{figure}

We have measured a precise phase difference between the two eclipses and found $\Delta \phi = 0.0907 \pm 0.0002$ cycles. We also measured the ratio of the widths of the eclipses, $w_p/w_s = 0.950 \pm 0.038$, where the subscripts $p$ and $s$ refer to the primary and secondary, respectively. From this we can make an initial estimate of the orbital eccentricity.  We use the expression for $\Delta \phi$ by Stern (1939):
\begin{equation}
\Delta \phi = \frac{1}{2}-\frac{1}{\pi}\left\{\arctan\left[\frac{e \cos \omega}{(1-e^2)^{1/2}}\right]+(1-e^2)^{1/2} \frac{e \cos \omega}{1-e^2 \sin^2 \omega} \right\}
\label{eqn:sterne}
\end{equation}
to solve numerically for the allowed value of eccentricity, $e$, for any given argument of periastron, $\omega$.  The results are shown in Figure~\ref{fig:e_P_contour}. From this plot we see that the minimum allowed value of $e$ is 0.705.  From the ratio of eclipse widths given above, and the approximate relation between that and $e \sin \omega$:
\begin{equation}
e \sin \omega \simeq \frac{1-w_p/w_s}{1+w_p/w_s}
\label{eqn:esinw}
\end{equation}
which, strictly speaking, only holds for small $e$ and inclination angles near $90^\circ$, we can say that $e \sin \omega \approx 0.026 \pm 0.020$ is at most a small number.  Thus, we expect $\omega$ to be within $\sim$5$^\circ$ of either 0$^\circ$ or 180$^\circ$.  This result, in combination with Figure~\ref{fig:e_P_contour} indicates that, in fact, $e=0.708^{+0.004}_{-0.001}$.

\subsection{Archival Data}
\label{sec:archival}

\begin{figure}
\null{} \hglue0.6cm \includegraphics[width=0.8\columnwidth]{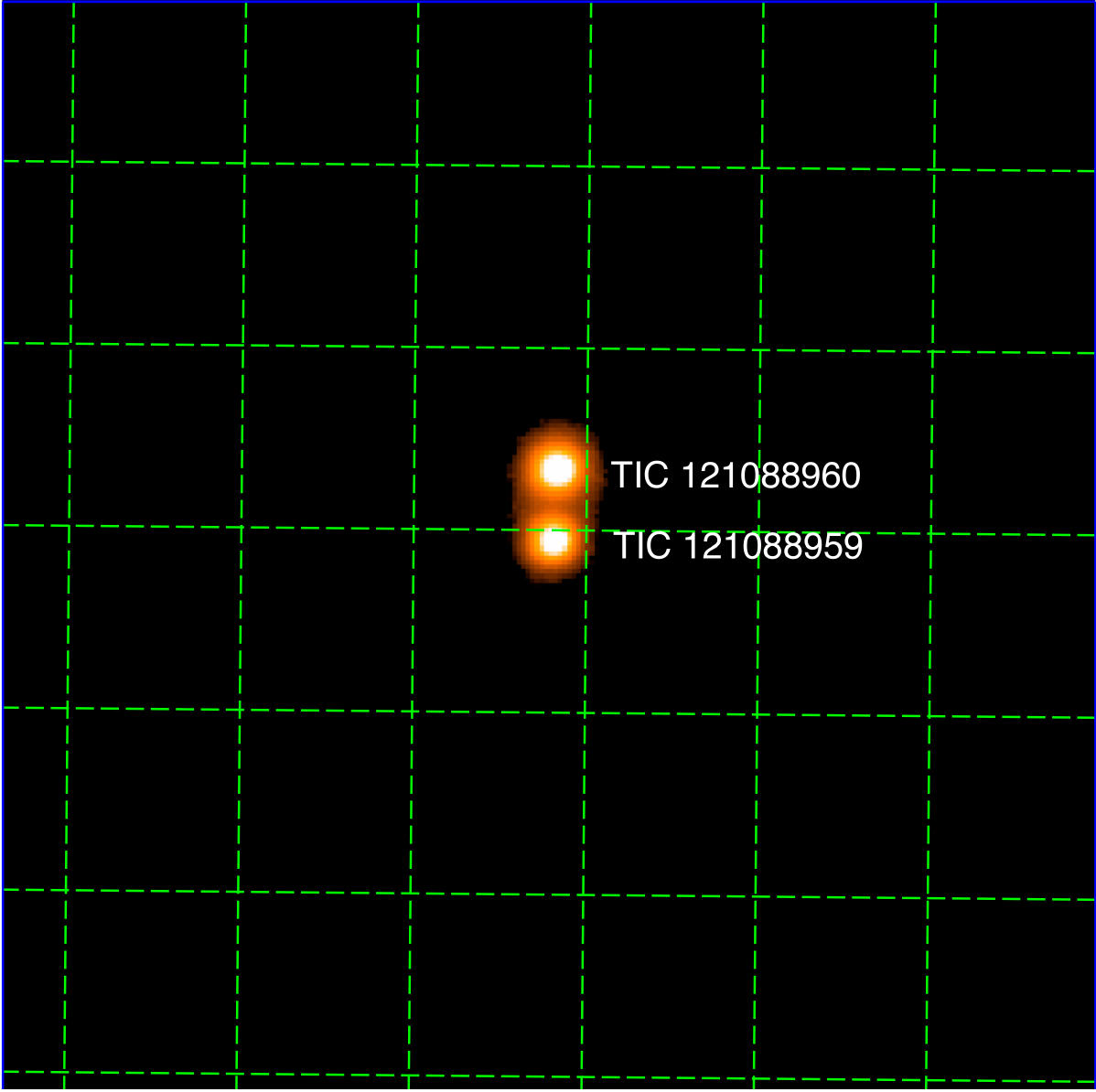}
\caption{PanSTARRS image of two bound M stars, TIC 121088959 and 121088960, the latter of which likely hosts the eccentric EB. The grid lines are spaced by $10'' \times 10''$.  The two images are separated by $3.9''$ on the sky.  The projected distance between the two images is $\sim$320 AU on the sky.}
\label{fig:PS_image}
\end{figure}

We have collected some of the important archival astrometric and magnitude values for the pair of stars in Table \ref{tab:mags}. The space motions determined by Gaia strongly indicate the two stars are gravitationally bound. However, the astrometric parameters of TIC 121088960 have larger uncertainties than TIC 121088959, so we checked the Renormalised Unit Weight Error (RUWE)\footnote{The RUWE parameter \citep{Lindegren2021a} provides an informative goodness-of-fit statistic. For values `significantly larger than 1.0 (say, $>1.4$) this could indicate that the source is non-single', or, in this case, not just a close binary.}. Gaia reports RUWE values of 1.08 and 4.27 for TIC 121088959 and TIC 121088960, respectively, with the latter value being abnormally large. Furthermore, for this same star, Gaia reports 0.75 for the astrometric\_excess\_noise parameter, and 213 for the astrometric\_excess\_noise\_sig parameter. Recent studies have shown that RUWE values $\gtrsim 1.24$ indicate the presence of unresolved companions \citep[e.g.][]{Rizzuto2018, Belokurov2020}. This, plus the large astrometric\_excess\_noise parameters, allow us to argue that TIC 121088960 is the host of unresolved stars, which likely includes the EB we are reporting on here.

\begin{table}
\centering
\caption{Properties of the TIC 121088959 and TIC 121088960 System}
\begin{tabular}{lcc}
\hline
\hline
Parameter & TIC 121088959 & TIC 121088960 \\
\hline
RA (J2000) (h m s)& 03:47:50.30 &  03:47:50.29 \\
Dec (J2000) ($^\circ \ ^\prime \ ^{\prime\prime}$) &  $-18$:54:11.65 & $-18$:54:07.78 \\
$T$$^a$ & $13.743 \pm 0.007$ & $13.343 \pm 0.006$ \\
$G$$^b$ & $14.994 \pm 0.001$ & $14.480 \pm 0.001$ \\
$G_{\rm BP}$$^b$ & $16.494 \pm 0.005$  & $16.072 \pm 0.003$ \\
$G_{\rm RP}$$^b$ & $13.793 \pm 0.001$ & $13.256 \pm 0.001$\\
$B^a$ & $17.026 \pm 0.112$ & $16.229 \pm 0.184$\\
$V^a$ & $15.305 \pm 0.067$ & $15.000 \pm 0.200$ \\
$J^c$ & $12.144 \pm 0.030$   & $11.531 \pm 0.051$ \\
$H^c$ & $11.546 \pm 0.032$  & $10.911 \pm 0.056$ \\
$K^c$ & $11.256 \pm 0.031$ & $10.701 \pm 0.055$ \\
W1$^d$ & ... & $9.993 \pm 0.024$ \\
W2$^d$ & ... & $9.814 \pm 0.021$ \\
W3$^d$ & ...  & $9.818 \pm 0.042$ \\
W4$^d$ & ... & $> 8.382$  \\
$T_{\rm eff}$ (K)$^d$ & $3828 \pm 157$ & $3888 \pm 200$ \\
Distance (pc)$^b$ & $ 83.67 \pm 0.39$ & $ 81.52 \pm 0.60$\\
$\mu_\alpha$ (mas ~${\rm yr}^{-1}$)$^b$ & $+10.33 \pm 0.03$ & $+8.77 \pm 0.08$ \\
$\mu_\delta$ (mas ~${\rm yr}^{-1}$)$^b$ &  $-46.91 \pm 0.02$ & $-41.99 \pm 0.08$ \\
\hline
\label{tab:mags}  
\end{tabular}

{\bf Notes.}  (a) ExoFOP (exofop.ipac.caltech.edu/tess/index.php).  (b) Gaia EDR3 (\citealt{Lindegren2021a};
\citealt{Lindegren2021b}; \citealt{GaiaEDR3}).  (c) 2MASS catalog \citep{Skrutskie2006}.  (d) WISE point source catalog \citep{Cutri2013}.
\end{table}

\subsection{Spectroscopic follow-up observations}
\subsubsection{iSHELL Observations}
\label{sec:iSHELL}

We observed TIC 121088959 and TIC 121088960 using iSHELL on NASA's InfraRed Telescope Facility (IRTF). iSHELL is a cross-dispersed near-infrared spectrograph covering a wavelength range of $\sim1.1 \mu m - 5.3 \mu m$. There are two slit options that yield resolving powers of $R=\lambda/\Delta\lambda = 35,000$ and $R=\lambda/\Delta\lambda = 75,000$. We used the K2 filter, covering from 2.09 $\mu m$ to 2.38 $\mu m$ and the $0\farcs75$ slit, resulting in the $R=\lambda/\Delta\lambda = 35,000$. TIC 121088959 was observed on the UT nights of 2021 February 9, 10, and 11, and TIC 121088960 was observed on the UT nights of 2021 February 9 and 11. On each night, we took calibration observations including dome flats and arc lamp for each science observation as required for iSHELL, followed by an A0V star observation. We reduced the spectra using the publicly available reduction pipeline for iSHELL, \texttt{Spextool} \citep[][]{spextool} and telluric corrected using \texttt{xtellcor}. Among the 29 orders in the K2-band data, we used 4 through 8, 11, and 15 for the analysis, which did not contain any obvious hot or bad pixels. We used the BT-Settl model spectra \citep[][]{Allard2012} as the RV templates and the models were obtained from the Spanish Virtual Observatory (SVO) website.\footnote{http://svo2.cab.inta-csic.es/theory/newov2/index.php} We used the 3300K and 3500K models, both with the solar metallicity and logg of 5.0. The effective temperatures were chosen as noted on the ExoFOP-TESS website.

To calculate the RVs, we followed the procedure described in \citet[][]{Han2019}. We first matched the BT-Settl models to have the same resolution as the iSHELL spectra. We interpolated both the science and the template spectra onto a logarithmic wavelength scale for a uniform sampling in velocity space. As mentioned in Section \ref{sec:discovery}, because we were not certain of whether the eclipses we see in the {\em TESS} data are from TIC 121088959 or TIC 121088960, we used the Two-dimensional CORrelation technique \citep[TODCOR,][]{TODCOR} to detect the SB2 RVs. We calculated the RV for each order and adopted the mean of the RVs as the measured RV. For the uncertainties, we took the standard deviation of the RVs across the orders divided by the square root of the number of orders used. Lastly, we applied the barycentric correction and report the five iSHELL RVs in Table \ref{tab:rv}.

\subsubsection{McDonald 2.7 m Coud{\'e}}
\label{sec:McDonald}

We observed TIC 121088960 using the \coude\ spectrograph on the McDonald Observatory, Harlan J. Smith 2.7 m telescope on the nights of 2021 January 30 and 31. From these observations we obtained the two additional RV points reported in Table \ref{tab:rv}. The Robert G.~Tull \coude\ is a cross-dispersed echelle spectrograph covering a wavelength range of 3400 to 10000 \AA\ with a resolution of $R\sim$60,000 using the 1\farcs2 slit \citep{Tulletal1995}. On both nights the seeing was sufficient to resolve the TIC 121088960--TIC 121088959 pair, and the slit orientation ensured minimal contribution from TIC 121088959. We took two 1500-s exposures on each night and reduced them using a custom python implementation of the standard reduction procedures. After bias and flat-field corrections, and cosmic ray rejection, the two echellograms are coadded before extraction to improve the signal-to-noise of the one-dimensional spectra. Wavelength solutions are derived from a series of ThAr comparison lamp spectra taken throughout the night. Our reduction results in detectable continuum emission in orders redward of 6400 \AA, achieving a peak signal-to-noise $\sim$10 in the reddest orders.

We measured RVs from the \coude\ spectra by computing spectral-line broadening functions \citep[BFs;][]{Rucinski1992,Tofflemireetal2019}. From a linear inversion of the observed spectrum with a narrow-lined template, the BF represents the average photospheric absorption-line profile as a function of velocity, which can be used to measure the stellar radial and rotational velocities. It can also determine the number of stellar components in the spectrum (i.e., ``double-lined'' systems). We tested a grid of narrow-lined synthetic templates from \citet{Husseretal2013} with 100 K steps in effective temperature, selecting the 3300 K template as the model producing the highest signal-to-noise, combined BF. The BF is computed for eight orders that contain sufficient signal and are free of telluric contamination (6400--8900\,\AA). The resultant BFs are combined first to determine the number of stellar components present, and then recombined, weighted by the noise in regions devoid of stellar components. We detect only one stellar component, which is fit with a rotationally-broadened absorption-line profile \citep{gray_book} to determine the RV and $v \sin i$. Uncertainties on the fitted parameters are determined with a boot-strap approach. 10$^5$ combined BFs are made from a random sampling with replacement of the 8 individual orders, which are fit individually. The 68\% interval of the output fit-parameter distribution is our quoted uncertainty. We do not measure a significant change in the RV between our two spectra, finding an average RV of $-5.8 \pm 0.3$ \kms\ and $v \sin i$ of $9 \pm1$ \kms.

\begin{table*}
\begin{center}
\caption{Measured radial velocities of TIC 121088959 and TIC 121088960}
\begin{tabular}{@{}l c c c c}
\hline
Target & BJD & RV (km\,s$^{-1}$) & $\sigma_{\rm RV}$ (km\,s$^{-1}$) & Instrument\\
\hline
TIC 121088959	&	2459254.7878997	&	2.08	&	0.61	&	iSHELL	\\
TIC 121088959	&	2459255.8000591	&	2.17	&	0.26	&	iSHELL	\\
TIC 121088959	&	2459256.7510484	&	2.36	&	0.28	&	iSHELL	\\
\hline
TIC 121088960	&	2459254.7310623	&	$-5.87$	&	0.44	&	iSHELL	\\
TIC 121088960	&	2459256.7958389	&	$-5.74$	&	0.50	&	iSHELL	\\
TIC 121088960   &   2459245.6190646 &  $ -5.8$    &   0.4     &   Coud{\'e}\\
TIC 121088960   &   2459246.5944387 &   $-5.9$    &   0.9     &   Coud{\'e} \\
\hline
\label{tab:rv}
\end{tabular}
\end{center}
\end{table*}

\section{Analysis and Results}
\label{sec:analysis}
\subsection{Lightcurve model and fit}

We modeled the {\em TESS} short- and long-cadence data from S4 and S31, following the description of \citet[][]{Han2017}. However, we used the Simple Aperture Photometry (SAP) flux instead of the Pre-search Data Conditioning SAP (PDCSAP) flux. The PDCSAP lightcurves of the targets that are produced by the TESS Science Processing Operations Center \citep[SPOC;][]{Jenkins2016} or the Quick-Look Pipeline \citep[QLP;][]{Huang2020} have been deblended assuming there are only two light-contributing stars in the photometric apertures. In fact, in this work we show that there are four contributing stars in the system.  Therefore we used the SAP data in our analysis which makes no assumptions about what is in the aperture.

The primary and the secondary eclipse depths in the SAP lightcurve are $\sim$3.3\% and $\sim$2.2\% deep. Moreover, both the S4 and S31 data do not exhibit eclipses other than the ones from the 3.04-d period binary. Therefore, we analyzed the {\em TESS} data considering that the 3.04-day eclipses could be from unseen companions of TIC 121088960 and the EB contributes only $\sim$10\% to the total system light. We used a publicly available modeling code for detached EBs, \textit{eb} \citep[][]{Irwin2011}. Among the 37 free parameters of the \textit{eb} model, we list 13 of interest in Table \ref{tab:modelparm}. The details of the rest of the model parameters can be found in \citet[][]{Irwin2011}.

\begin{table}
\begin{center}
\caption{Modeling Parameters}
\begin{tabular}{@{}l c}
\hline
Parameter & Description \\
\hline
$J$ & Central surface brightness ratio (secondary/primary)\\
$(R_1 + R_2)/a$ & Fractional sum of the radii over the semi-major axis\\
$R_2/R_1$ & Radii ratio\\
$\cos{i}$ & Cosine of orbital inclination \\
$P$ (days) & Orbital period in days \\
$T_0$ (BJD) & Primary mid-eclipse time \\
$e \cos{\omega}$ & Orbital eccentricity $\times$ cosine of argument of periastron  \\
$e \sin{\omega}$ & Orbital eccentricity $\times$ sine of argument of periastron\\
$L3$ & Third light contribution\\
LDLIN1 & Linear limb-darkening coefficient for the primary\\
LDNON1 & Square root limb-darkening coefficient for the primary\\
LDLIN2 & Linear root limb-darkening coefficient for the secondary\\
LDNON2 & Square root limb-darkening coefficient for the secondary\\
\hline
\label{tab:modelparm}
\end{tabular}
\end{center}

\end{table}

In all modeling procedures, we first searched for the best-fit model by $\chi^2$ minimization using the Levenberg-Marquardt technique \citep[\texttt{mpfit}, ][]{Markwardt2009}. We then employed the Markov chain Monte Carlo (MCMC) algorithm using \texttt{emcee} \citep[][]{Foreman-Mackey2013}, explored the parameter spaces, refined the model, and determined the uncertainties. The parameters of the best-fit model from \texttt{mpfit} were used as the starting parameters of the MCMC walkers. For all our MCMC runs, we used 300 walkers, each with 15000 steps and uniform priors for all 13 model parameters. Moreover, we excluded the majority of the out-of-eclipse fluxes in modeling and used only the region covering from orbital phases $-0.08$ to +0.07 (see Figure~\ref{fig:fold}), to save computational time.

As done in \citet[][]{Han2017} and \citet[][]{Han2019}, we made two modifications in modeling using \texttt{eb}. The first is with $e\cos{\omega}$ and $e\sin{\omega}$ where we stepped in $\sqrt{e} \cos{\omega}$ and $\sqrt{e} \sin{\omega}$ and converted them to  $e\cos{\omega}$ and $e\sin{\omega}$ for the model computation. This is to ensure that the uniform priors of $e\cos{\omega}$ and $e\sin{\omega}$ do not bias toward high values of eccentricity \citep[][]{Ford2006}. The second is with the limb-darkening parameters where we converted the square-root limb-darkening coefficients of \texttt{eb} to the $q$'s defined by \citet{Kipping2013}. We stepped in the $q$'s, allowing them to vary only between 0 and 1, and converted the $q$'s back to the square-root limb darkening coefficients for the model computations. This is to ensure all possible combinations of $q$'s to be physical.

\begin{figure*}
\includegraphics[width=0.8\linewidth]{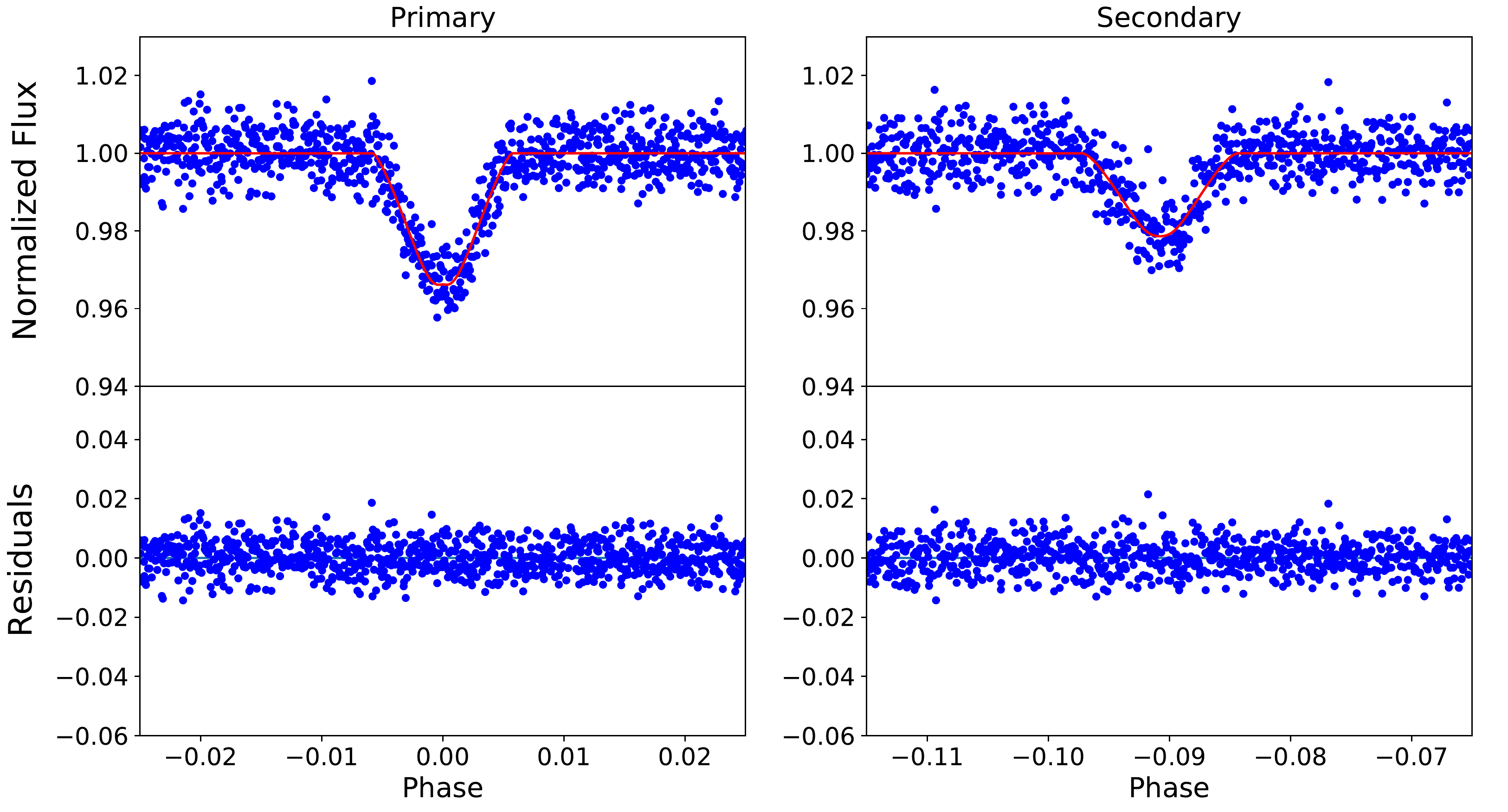}
\caption{Best-fit lightcurve model for the unseen EB. The upper panels show the short-cadence data in blue dots and the best-fit model as a solid red line. The lower panels show the residuals of the fit. Here we have defined phase zero of the lightcurve to be at the time of primary eclipse, as opposed to Figure 2 where phase zero was defined to be at periastron for aesthetic purposes.}
\label{fig:bestfit}
\end{figure*}

\begin{figure*}
\includegraphics[width=0.8\linewidth]{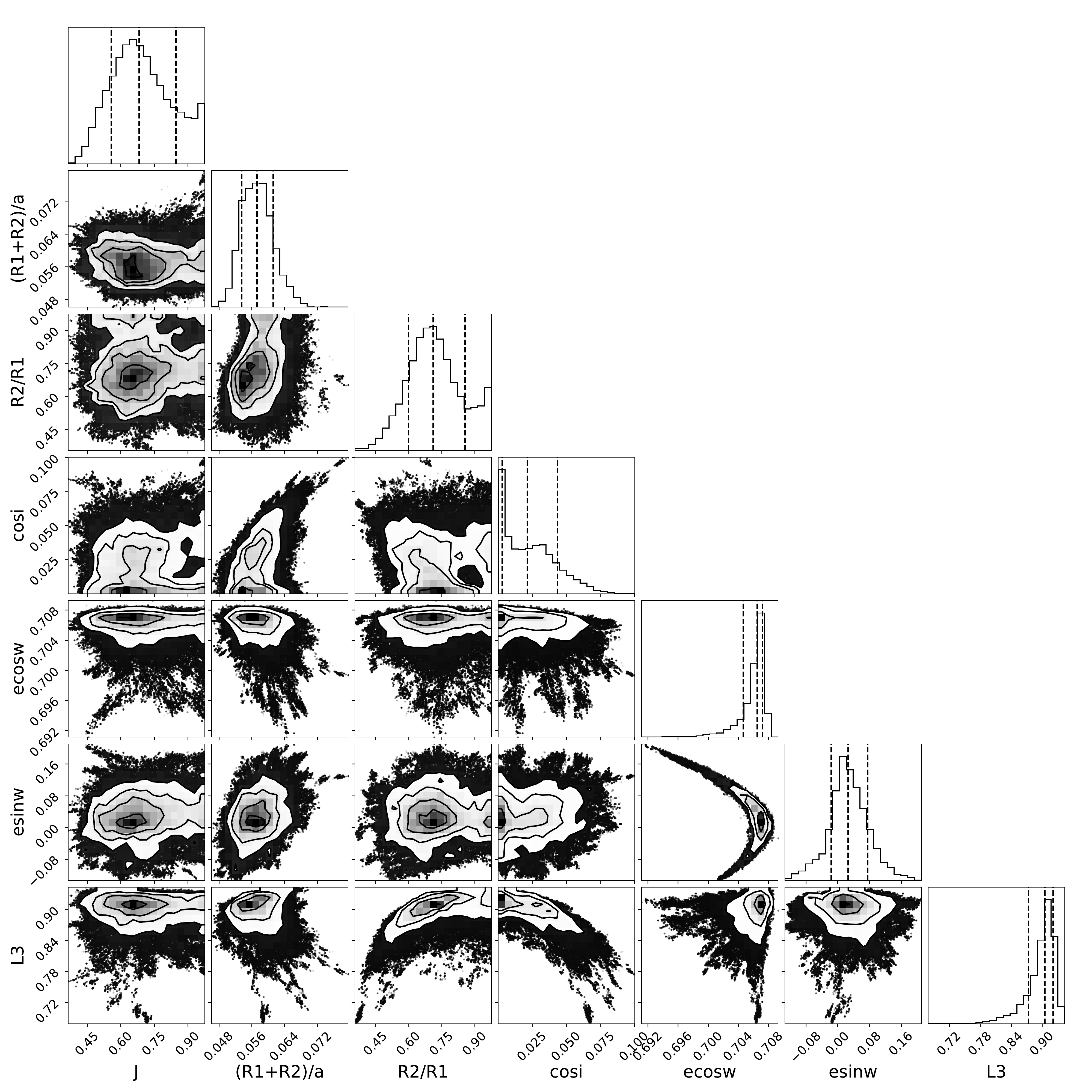}
\caption{Triangle plot from the lightcurve fit of S31 short-cadence data showing the posterior distributions of the extracted stellar parameters. The dashed lines in the histogram mark the 16$^{th}$, 50$^{th}$, and 84$^{th}$ percentiles of the distributions. }
\label{fig:corner}
\end{figure*}

Once each MCMC run is finished, we visually inspected the walkers to ensure convergence. For the analysis, we removed the first 5000 steps of each walker as the "burn-in" and searched for the most probable model from a single step that has the highest likelihood. We report the set of parameters from the most probable model as our best-fit parameters instead of the median values of the posterior distributions. For the symmetric posterior distributions, we report the standard deviations as uncertainties. For the asymmetric posterior distributions, we calculate the 34.1$^{th}$ percentile around the maximum likelihood, take the difference between the values of the maximum likelihood and the 34.1$^{th}$ percentile values and report them as asymmetric uncertainties. Table \ref{tab:fitted_parameters} contains the best-fit parameters extracted from the short-cadence data. Figure \ref{fig:bestfit} shows the best-fit model as a red solid line (upper panels) and the residuals (lower panels) around the primary and the secondary eclipses. Figure \ref{fig:corner} shows the posterior distributions of the extracted stellar parameters. The dashed lines in the histogram mark the 16$^{th}$, 50$^{th}$, and 84$^{th}$ percentiles of the distributions. We assumed that the primary star is hotter and larger than the secondary star and therefore limited the surface brightness ratio and the radii ratio to not exceed 1.

We note that the derived EB parameters are not (yet) in physical units until we are able to determine the masses of the component stars of the EB and their semi-major axis via supplemental information such as SED fitting (see Sect.~\ref{sec:PhotConstraints}). We also inspected to see if there are any eclipse timing variations (ETV) that are induced by TIC 121088960. EBs can experience changes in the orbital period from the classical R\o{}mer delays or dynamical delays due to the presence of a third body. Both effects could provide useful information on the masses of the third body and the triple system. We phase folded S4 and S31 data from {\em TESS} using the best-fit $T_0$ and period in Table \ref{tab:fitted_parameters}. We found no divergence in either the primary or the secondary eclipse and hence no signs of ETVs over the 2-year baseline.

To solve for the orbital eccentricity ($e$) and the argument of periastron ($\omega$) of the EB, we took the posterior distributions of $e \cos{\omega}$ and $e \sin{\omega}$, which result in $e$ = 0.7085$^{+0.0013}_{-0.0006}$ and $\omega$ = 4.85$^{+3.05^\circ}_{-1.82^\circ}$. These values are consistent with the simple approach of using equations \ref{eqn:sterne} and \ref{eqn:esinw}. To further investigate the orbital elements, we compared $e \cos{\omega}$ and $e \sin{\omega}$ from the S31 short-cadence data with those from the S4 long-cadence data. From the S4 long-cadence data, we extracted $e \cos{\omega} = 0.7097^{+0.0069}_{-0.0019}$ and $e \sin{\omega} = 0.0016^{+0.0583}_{-0.0781}$, which result in $e$ = 0.7097$^{+0.0017}_{-0.0014}$ and $\omega$ =  0.13$^{+4.72^\circ}_{-6.28^\circ}$. As expected, $e$ and $\omega$ from the S4 long-cadence data have larger uncertainties but are consistent with those from the S31 short-cadence data within the measurement uncertainties. This supports the fact that we did not measure any apsidal motion over the two-year baseline.

\begin{table}
\begin{center}
\caption{Extracted parameters of the unseen EB from S31 short-cadence data}
\begin{tabular}{l c c}
\hline
Fitted in Lightcurve Analysis & Primary & Secondary\\
\hline
$J$ & \multicolumn{2}{c}{0.688 $^{+0.066}_{-0.054}$ } \\
$(R_1 + R_2)/a$ & \multicolumn{2}{c}{0.0553 $\pm$ 0.018 } \\
$R_2/R_1$ & \multicolumn{2}{c}{0.676 $^{+0.048}_{-0.054}$ } \\
$\cos{i}$ & \multicolumn{2}{c}{0.022 $\pm$ 0.010} \\
$P$ (days) & \multicolumn{2}{c}{3.04358 $\pm$ 0.00001} \\
$T_0$ (BJD) & \multicolumn{2}{c}{2458825.0657 $\pm$ 0.0005} \\
$e \cos{\omega}$ & \multicolumn{2}{c}{0.7059 $^{+0.0006}_{-0.0011}$} \\
$e \sin{\omega}$ & \multicolumn{2}{c}{0.0599 $^{+0.0378}_{-0.0224}$} \\
$L3$ & \multicolumn{2}{c}{0.899 $^{+0.009}_{-0.014}$}\\
LDLIN & -0.053 $\pm$ 0.353 & -0.637 $\pm$ 0.348 \\
LDNON & -0.748 $\pm$ 0.656 & 1.347 $\pm$ 0.598  \\
\hline
Calculated orbital parameters & Primary & Secondary\\
\hline
$R/a$ & 0.0330 $^{+0.0011}_{-0.0010}$ & 0.0223 $\pm$ 0.0015  \\
$i$ ($^\circ$) & \multicolumn{2}{c}{ 88.74 $\pm$ 0.69}  \\
$e$ & \multicolumn{2}{c}{0.7085 $^{+0.0013}_{-0.0006}$}\\
$\omega$ ($^\circ$) & \multicolumn{2}{c}{4.85 $^{+3.05}_{-1.82}$} \\
\hline
\label{tab:fitted_parameters}
\end{tabular}
\end{center}
\end{table}

\subsection{Radial velocity data}

As shown in Table \ref{tab:rv}, we did not detect the secondary component of either TIC 121088959 or TIC 121088960 in any of the iSHELL or Coud\'e spectra. Our data span an 11-day baseline and the RVs of each star remain consistent to within the uncertainties over the 11 nights with a variation in RV of less than 0.5 \kms. Furthermore, the measured RV difference between TIC 121088959 and TIC 121088960 is consistent at $\sim$8 \kms. Considering that the two stars are gravitationally bound, as evidenced by how similar their parallaxes and PMs are, this RV difference is large. We attribute the relatively large RV difference to the unseen EB that exerts a gravitational pull on TIC 121088960. Details on the orbital motion of the triple is discussed in Section \ref{sec:tripleorbit}. Future RV follow-up observations allowing longer baselines are required to detect any changes in the orbital motion.

\section{Which Star Hosts the Eccentric Binary?}
\label{sec:host}
Here we establish that only TIC 121088959 or TIC 121088960 can be the host of the eccentric EB with the far higher probability being assigned to TIC 121088960.

\begin{figure}
\includegraphics[width=0.99\linewidth]{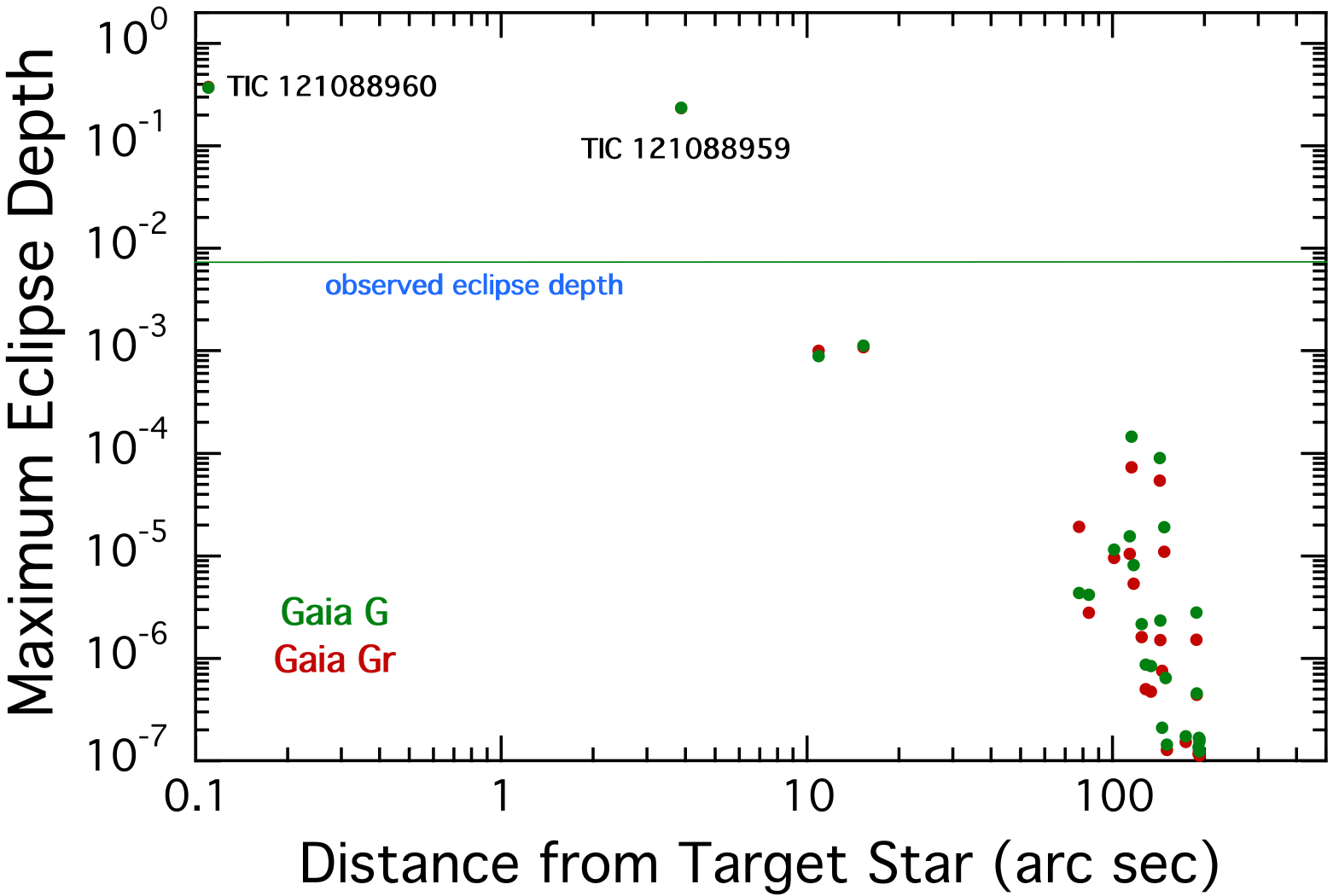}
\caption{All 25 neighbor stars from Gaia out to 200$''$ from TIC 121088960.  The plot shows the maximum possible eclipse depths attributable to those stars in G (green) and G$_R$ (red) bands and takes into account both the magnitude and the distance from the target star.}
\label{fig:neighbors}
\end{figure}

We start by showing in Figure~\ref{fig:neighbors} the maximum possible eclipse depths for all twenty five Gaia stars within 200$''$ of the target star vs. their actual distance from TIC 121088960.  Here we have taken into account both the magnitude of the star and the decay of the optical point spread function of the {\em TESS} camera.  The case for no stellar leakage into the photometric aperture is aided by the fact that all stars within 90$''$ of the target have $G > 20$.  Thus, from this perspective, none of the neighbor stars seems capable of introducing a significant, but spurious, eclipsing signal into the time series for the target stars.

\begin{figure*}
\includegraphics[width=0.3\linewidth]{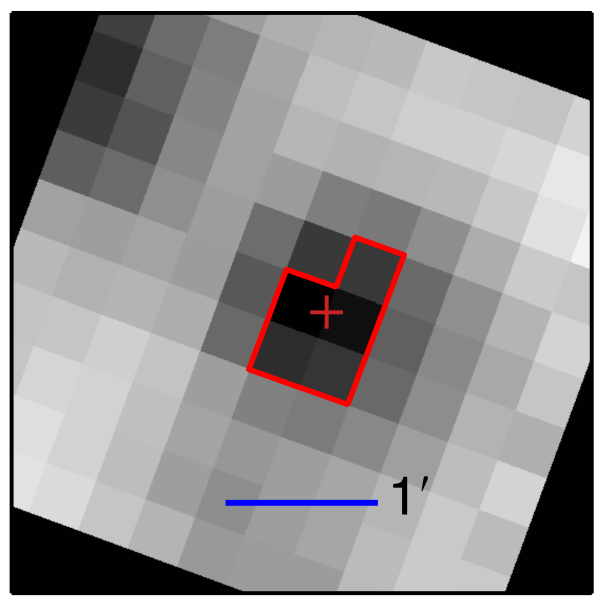}
\includegraphics[width=0.3\linewidth]{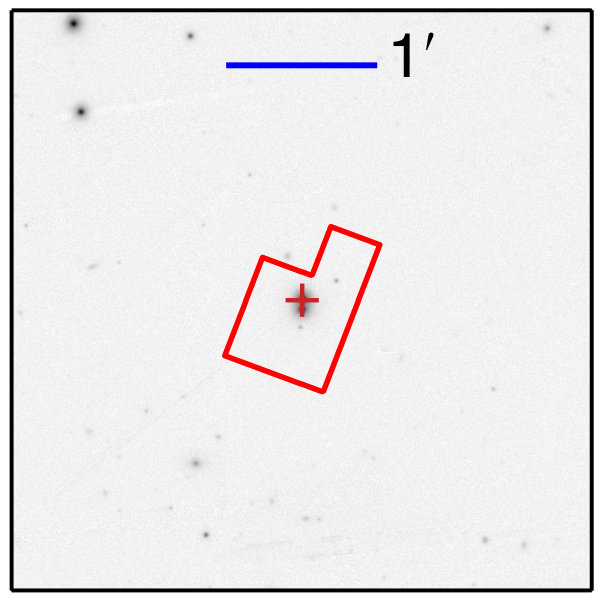} \hglue0.03cm
\includegraphics[width=0.3\linewidth]{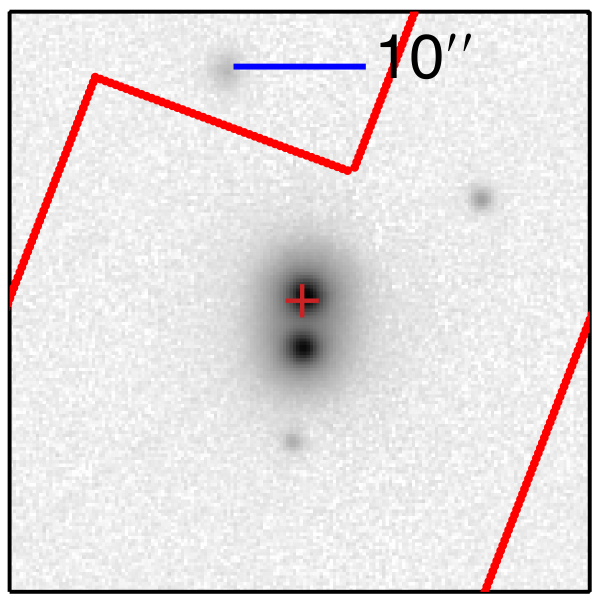}
\caption{{\em TESS} pixel level detail for TIC 121088959 and TIC 121088960.  {\em Left panel:} An actual {\em TESS} image with the photometric aperture superposed. {\em Middle panel:} Photometric aperture superposed on the Pan-STARRS image (see also Figure~\ref{fig:PS_image}). {\em Right panel:} Same as middle panel but zoomed in.}
\label{fig:pixel_level}
\end{figure*}

\begin{figure}
\includegraphics[width=0.955\linewidth]{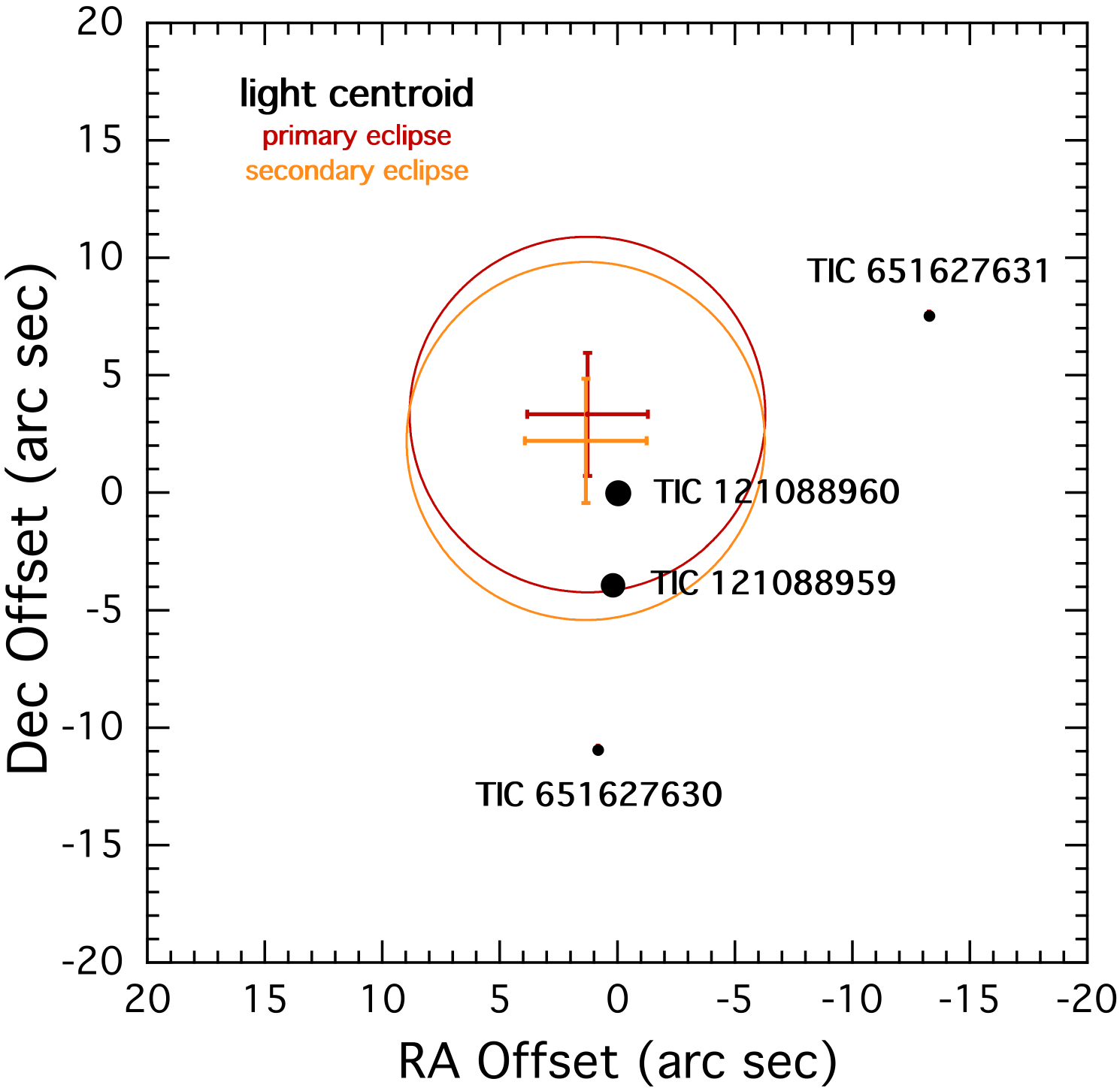}
\caption{Light centroid for the primary eclipse (red cross) and secondary eclipse (orange cross).  The locations of TIC 121088960 and TIC 121088959 are indicated as heavy filled circles on the plot.  The arms on the crosses represent the $\pm 1-\sigma$ uncertainties on the light centroids, while the circles correspond to the $3-\sigma$ range on the centroid locations. These demonstrate that the eclipses occur with a few arc seconds of the target stars.  Specifically, the eclipses are distinctly closer to TIC 121088960 and rule against TIC 121088959 at the 2.9 $\sigma$ and 2.4 $\sigma$ level for the primary and secondary eclipses, respectively. The other two marked stars are at $>20$\,th magnitude.}
\label{fig:centroids}
\end{figure}

We can also utilize the position of the light centroid of the difference image (between in and out of eclipse) to evaluate quantitively where the eclipses are located on the sky.  Figure \ref{fig:pixel_level} (left panel) shows the actual {\em TESS} image of the two target stars with the photometric aperture superposed.  The middle and right panels of Figure~\ref{fig:pixel_level} show the photometric aperture superposed on the Pan-STARRS image using two different zoom values.  Clearly the two target stars are not resolved by the large pixel size and photometric aperture.  However, we can measure the light centroid fairly accurately both in and out of eclipse.  The results are shown in Figure~\ref{fig:centroids}.  The top panel gives the light centroid of the difference image (in and out of eclipse) for the primary, while the bottom panel is the same for the secondary eclipse.  Clearly the eclipse light centroid matches TIC 121088960 better than it does TIC 121088959, at the 3.0 $\sigma$ and 2.5 $\sigma$ levels for the primary and secondary eclipses, respectively. From this exercise we conclude that (i) the eclipses occur within 7$''$ of the target stars, and, (ii) within the measurement uncertainty, quite near to TIC 121088960, but 3 $\sigma$ away from TIC 121088959.

It is, of course, possible that the EB is a random background object that happens to lie very near one of the target stars, but cannot be resolved by Gaia.  Such a star would have to have at least $G = 17$ in order to produce a 3.5\% and a 2\% eclipse in the presence of the two target stars. We find 5 stars with $G < 17$ in an area covering $1.3 \times 10^5$ arc-sec$^2$ and centered on the target stars. We estimate the combined area around the two target stars where Gaia might be `blind' to another star as $\sim$3 arc-sec$^2$. From this we estimate that there is only a $\sim10^{-4}$ probability of finding such an unrelated EB lying accidentally near the target stars,  and therefore rather unlikely.

In summary, we conclude that the eccentric EB (i) is definitely hosted by either TIC 121088960 or TIC 121088959, but (ii) is by far most likely associated with TIC 121088960.

\section{Motivation for a 2+1+1 Quadruple System}
\label{sec:motivation}
A number of lines of evidence have led us to the conclusion that TIC 121088959 and TIC 121088960 (see Figure~\ref{fig:PS_image}) form a quadruple system with the eccentric EB -- an unresolved part of TIC 121088960.  In Figure~\ref{fig:schematic}, we show how we envision the structure of this quadruple system, and then explain how we arrived at this conclusion. We use conventional nomenclature for the stars in a hierarchical multi-stellar system.  Thus, TIC 121088959 becomes `star B', while TIC 121088960 is composed of a `star A' plus the eccentric binary, stars `Ca' and `Cb'.

Here we summarize the evidence in favor of this being a quadruple system (A+C+B) rather than simply an eccentric binary (A) plus a distant companion (B): \\
  \indent (i) The two eclipse depths are quite small at only 3.3\% and 2.2\% for the primary and secondary eclipses, respectively.
  This suggests that either the eclipses are highly grazing, or the light from the EB is being considerably diluted by brighter stars (A and B).\\
   \indent (ii) RVs (three for TIC 121088959 and four for TIC 121088960) measured over an 11-day interval show no evidence for any changes at the $\sim$0.5 km s$^{-1}$ level (see Table 2). The easiest way to explain this lack of RV change is that there is another brighter star, other than the EB, in one of the images.  Otherwise, the expected RVs of $\sim$15 km s$^{-1}$ even in the apastron region would have been detected.  We carefully checked the orbital phases of the RV measurements, and if the EB were fully represented by either TIC 121088959 or TIC 121088960, the RV changes could not have escaped detection regardless of which one hosts the EB.\\
   \indent (iii) In Sect.~\ref{sec:host} we showed at the $\sim$3-$\sigma$ level that the EB is most likely associated with TIC 121088960. Thus, the natural explanation for the lack of RV changes would be explained by the fact that star A is diluting the light from the EB (C) to the point where the RVs are locking onto star A and the lines from the EB (C) are not being detected at all.  \\
   \indent (iv) The difference in RVs between TIC 121088959 and TIC 121088960 according to our measurements is $8 \pm 0.3$ km s$^{-1}$ (see Table 2). This is too large to be accounted for by orbital motion between TIC 121088959 and TIC 121088960.  These two stellar images are $3\farcs9$ apart which corresponds to a projected physical separation $\sim$320 AU.  If we adopt trial masses for these two objects of $\sim$0.4 M$_\odot$, that yields a characteristic orbital period of 6000 years, assuming an approximately circular orbit.  The corresponding relative orbital speeds would be only 1.5 km s$^{-1}$.  This is much too low to explain the observed 8 km s$^{-1}$ difference in RVs.  The much more likely explanation is that these speeds result from the EB (C) pulling the dominant star in TIC 121088960 (A) around in a much shorter period orbit of years. \\
    \indent (v) The difference in proper motions (PMs) between TIC 121088959 and TIC 121088960 measured by Gaia over a nearly 3-year interval is $5 \pm 0.1$ mas yr$^{-1}$, or 2.0 km s$^{-1}$ at the distance of these two stars. By the same argument used in point (iv) this relative speed on the sky is on the high side to be accounted for by orbital motion between TIC 121088959 and TIC 121088960. \\
    \indent (vi) TIC 121088960 has an elevated RUWE (4.3) and significant astrometric\_excess\_noise, strongly hinting that it has multiple stellar components. \\
    \indent (vii) Both TIC 121088959 and TIC 121088960 are at distances of close to 83 pc with a formal radial distance separation of $3.3 \pm 0.9$ pc. We take the similarities in RV, PM, and distance to indicate that these are a gravitationally bound pair. \\

 Thus, we find that the preponderance of evidence points toward a quadruple system with the EB (hereafter called binary `C') being a fainter member of TIC 121088960, where the brighter star in that image is designated as star `A'.  TIC 121088959 is, as far as we know, a single star which we label as `B', and is bound to the AC triple subsystem.

\section{Estimating the Parameters of the Inner Triple System AC}
\label{sec:tripleorbit}

\begin{figure}
\centering
\includegraphics[width=0.45\textwidth]{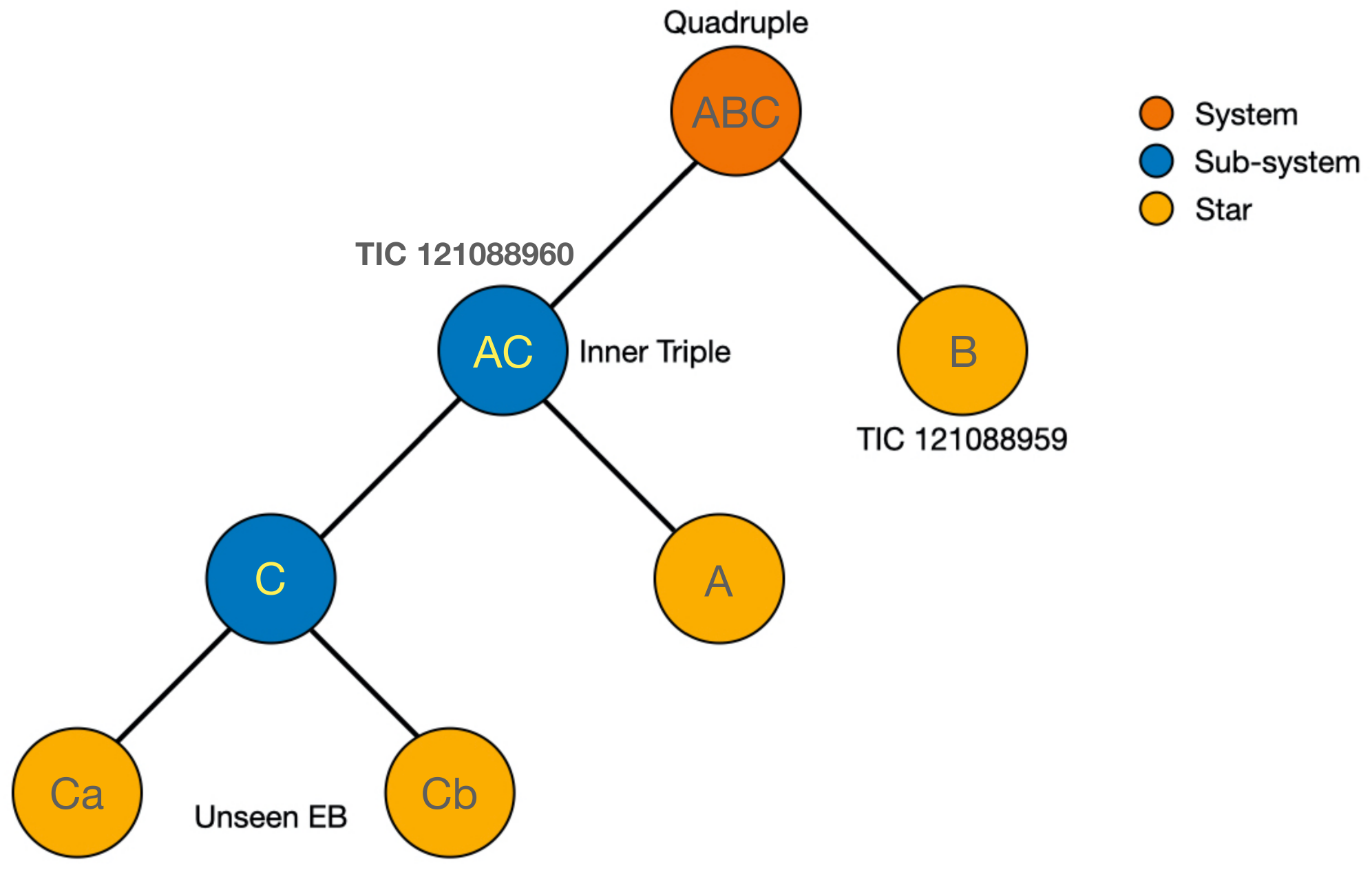} \vglue0.5cm
\includegraphics[width=0.45\textwidth]{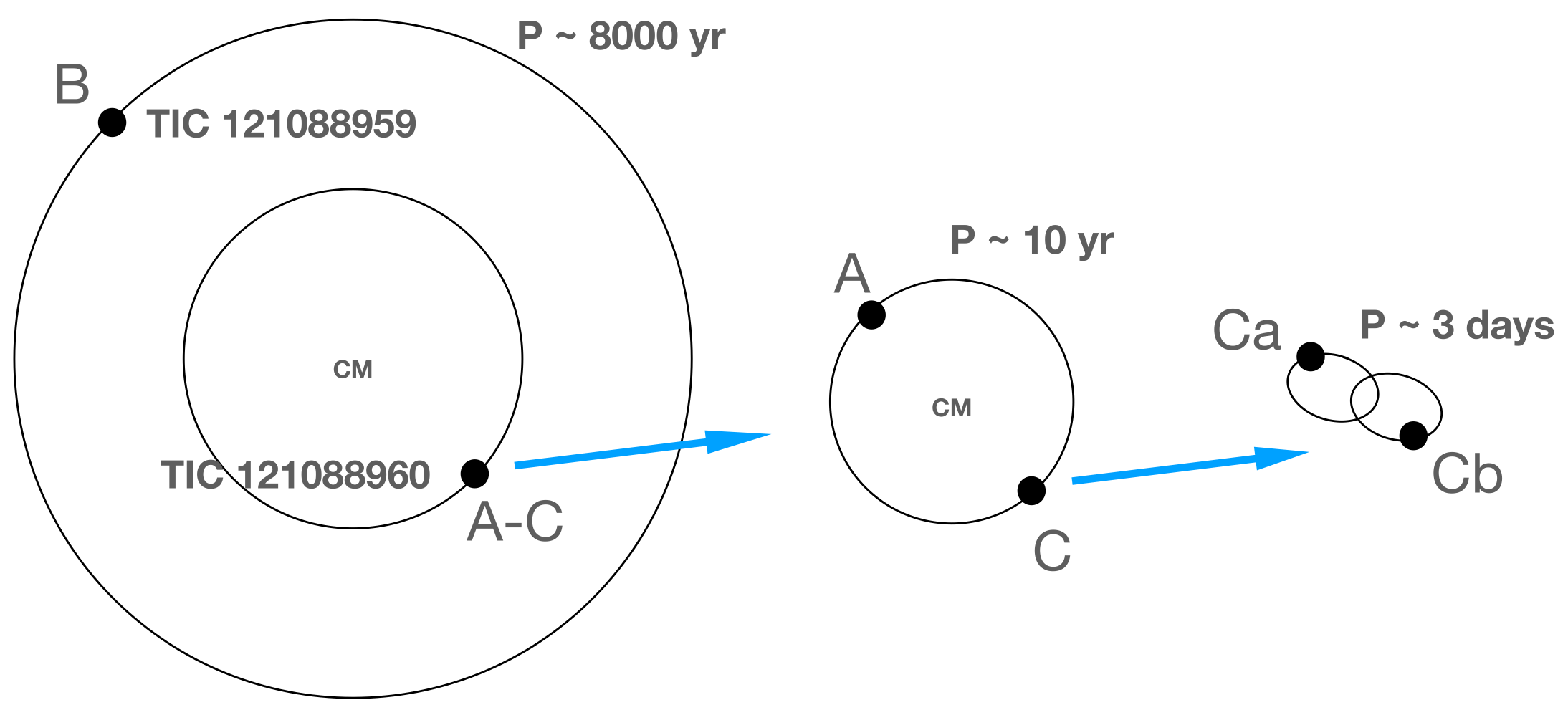}
\caption{Schematic of the 2+1+1 quadruple system TIC 121088959/60. {\em Top panel:} organizational tree.  {\em Bottom panel:} sketch of the orbits from above an assumed common orbital plane.  The subsystems are shown broken out separately since the range of the three scales is $\sim$$10000:100:1$, and would be illegible if all were superposed.}
\label{fig:schematic}
\end{figure}

Here we first estimate the constituent masses of the triple AC (as well as that of star B in the same process) using a fit to the spectral energy distribution (SED).  We then use dynamical considerations to constrain the orbit of the inner triple (AC; see Figure~\ref{fig:schematic}), including its period.  Less can be said about the outer quadruple orbit (`ABC'), but we provide some estimates of the outer period.

\subsection{Photometric Constraints---SED Fitting for the Masses}
\label{sec:PhotConstraints}

\begin{figure}
\centering
\includegraphics[width=1.0\linewidth]{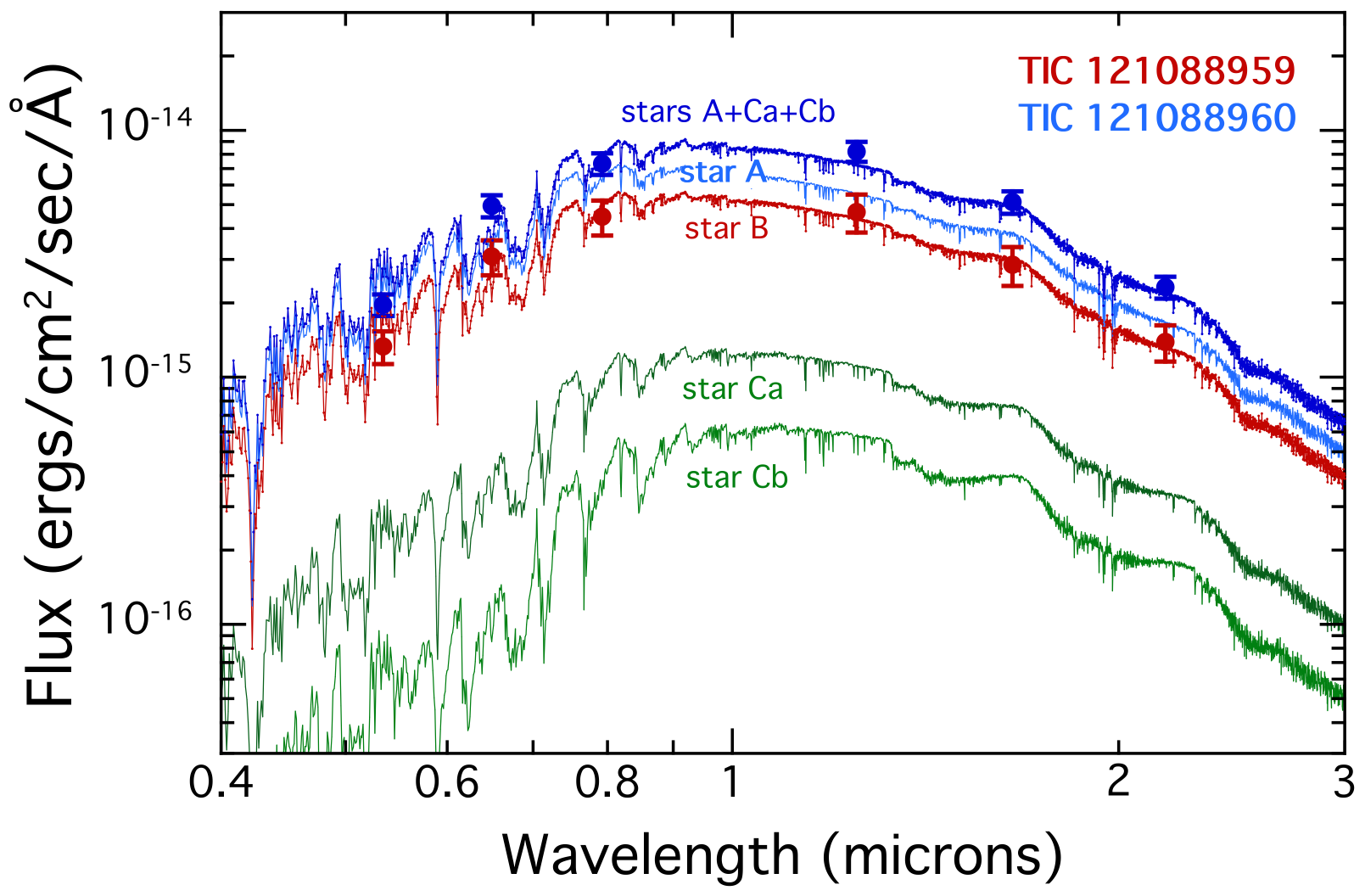}
\caption{SED fits to the quadruple system TIC 121088959/60.  The red curve is the model for star B (TIC 121088959), while the blue curve is for the composite of stars Ca, Cb, and A. The model SED for the individual stars Ca, Cb, and A are shown in green and light blue, respectively. The match to the measured fluxes is made by integrating the model fluxes over the appropriate filter bands.}
\label{fig:SED}
\end{figure}

We collected the available spectral energy distribution (SED) points for the two stellar images from VizieR \citep[][]{vizier}.  There are three SED points from Gaia \citep[][]{GaiaEDR3} and three from 2MASS \citep[][]{Cutri2003}.  The stars are clearly detected by WISE \citep[][]{Cutri2013} in their bands 1, 2, and 3, but unfortunately they are unresolved at these longer wavelengths (with resolutions of $\sim$$6\farcs1$, $6\farcs4$, and $6\farcs5$, respectively). So as to avoid ambiguities, we fit just the six SED points for each image where the two are clearly resolved. The SED measurements that we have used are shown in Figure~\ref{fig:SED} for each of the two images.

To fit the SEDs, our approach is as follows.  We use an MCMC approach where the four fitted parameters are the four stellar masses.  Because these stars are of quite low mass (all $< 0.5 M_\odot$, as we shall demonstrate), we assume that they are all firmly on the zero-age main sequence (ZAMS).  In that case we use mass-radius-temperature relations for ZAMS stars (see, in particular, Eqns.~(A1) and (A2) in \citealt{Rappaport2017}). The chosen masses in each link of the MCMC chain then yield the corresponding radii and values of $T_{\rm eff}$. For the spectral fits we used the stellar atmospheres models of BT-Settl \citep[][]{Allard2012}.

While we are fitting for only four free parameters (the 4 stellar masses) and there are 12 SED points, experience has shown that this is still insufficient to determine all four masses uniquely (see, e.g., \citealt{Powell2021}, \citealt{borkovits2021}, and \citealt{Kostov2021} for details). Therefore, in addition to the six SED points for each stellar image, we make use of four other important constraints.  From our fit of the EB lightcurve we take (i) the ratio of stellar radii in the EB to be $R_{\rm Cb}/R_{\rm Ca} = 0.676 \pm 0.1$; (ii) the sum of the scaled stellar radii to be  $(R_{\rm Cb}+R_{\rm Ca})/a = 0.055 \pm 0.018$; (iii) the eclipse depth ratio of primary to secondary to be $1.5 \pm 0.15$; and (iv) the third light for the eccentric EB to be $89.9\% \pm 1.5\%$.  These 16 total constrains, plus the assumption that the stars are on the ZAMS, then prove sufficient to fit uniquely for the four stellar masses\footnote{The approach of using SED fitting on multiple star systems has been demonstrated by numerous groups, but perhaps none more dramatically than for the sextuple star system TIC 168789840 (see \citealt{Powell2021}).}. The results are shown in Figure~\ref{fig:SED} and the fitted stellar parameters are given in Table \ref{tbl:MRTL}.

We find that the masses of the stars in the eccentric EB are $M_{\rm Ca} \simeq 0.190 \,{\rm M}_\odot$ and $M_{\rm Cb} \simeq 0.138 \,{\rm M}_\odot$.  The mass of star B (TIC 121088959) is $M_{\rm B} \simeq 0.386 \,{\rm M}_\odot$, while the brighter member of TIC 121088960 (star A) is only slightly more massive at $M_{\rm A} \simeq 0.430 \,{\rm M}_\odot$.

\begin{table*}
\centering
\caption{Mass Estimates for the Four Stars in TIC 121088959/60}
\begin{tabular}{lcccc}
\hline
\hline
Parameter    &   Star Ca & Star Cb & Star A & Star B  \\
\hline
 Mass (M$_\odot$) & $0.190\pm0.009$ & $0.138\pm 0.005$  & $0.430\pm0.008$  & $0.386\pm0.008$ \\
 Radius (R$_\odot$)  & $0.201\pm0.007$  & $0.160\pm0.005$   & $0.394\pm0.005$  & $0.358\pm0.007$ \\
 $T_{\rm eff}$ (K)  & $3228\pm25$  & $3033\pm30$   & $3512\pm10$  &  $3468\pm10$ \\
 Luminosity  (L$_\odot$) & $0.0040\pm0.0004$ &  $0.0020\pm0.0002$  & $0.0213\pm0.0007$ & $0.0167\pm0.0007$ \\
 a (R$_\odot$)  & $6.08 \pm 0.07$ & $6.08 \pm 0.07$ & ... & ... \\
\hline
\label{tbl:MRTL}
\end{tabular}
\end{table*}

\subsection{Constraining the Triple Orbit AC}
\label{sec:triple_orbit}

Armed with reasonably good estimates for the masses of the four stars comprising this system, we can now place some constraints on the orbit of the eccentric EB around what we have called ``star A''.  Let us refer to this orbit as the `AC' orbit or that of the `inner triple'.  There are five constraints that we use to limit the range of orbital periods and other parameters for the AC system.

These include the facts that (i) the AC system must be dynamically stable; (ii) the ABC system (i.e., that of the quadruple), must also be dynamically stable; (iii) there is no observed apsidal motion of the binary C ($2-\sigma$ limit of $\lesssim 10^\circ$ between the two {\em TESS} observations spaced by two years); (iv) the difference in RV between TIC 121088959 and TIC 121088960 (i.e., between stars A and B) is $8 \pm 0.3$ km s$^{-1}$ over an 11-day interval; and (v) there is a difference in PM between stars A and B of $5 \pm 0.2$ mas yr$^{-1}$.

The last two of these constraints (the difference in RVs and PMs) were from different epochs separated by a few years. Nonetheless, for simplicity in the calculations, we take this difference in time to be considerably less than the orbital period of the AC binary, and therefore these two constraints are applied locally around the orbits being examined.  This becomes a good approximation for $P_{\rm AC} \gtrsim 10$ years. Furthermore, with regard to these same two constraints, we assume that both the Gaia and ground-based observations are dominated by star A in the AC system (TIC 121088960). The expression for the minimum $P_{\rm AC}$ allowed before apsidal advance in the C binary would be observed is from \citet{Rappaport2017}, Eqn.~(13).  The lower and upper bounds on $P_{\rm AC}$ for the dynamical stability of orbits AC and ABC, respectively, are given by Eqn.~(16) of \citet{Rappaport2013}\footnote{See \citet{Mardling2001} and \citet{Mikkola2008} for the original expressions.} modified with a leading factor of $(1+e_{\rm inner})^{3/2}$ which we adopt from the Eggleton-Kiseleva stability requirement (\citealt{Eggleton1995}, Eqn.~1; \citealt{Mikkola2008}, Eqn.~10).  We take this to account for the fact that the more eccentric the inner binary is in a triple system, at a fixed binary orbital period, the more the two stars are separated at apastron.  Finally, with regard to the constraint on the stability of the ABC orbit we estimate a limit on the outer orbital period by noting that the projected physical separation of  AC and B on the sky is $\simeq 320$ AU.  We could use this as a proxy for the semimajor axis of the outer orbit.  However, to be somewhat more conservative, we take the semi-major axis of the outer orbit to be $\lesssim 1000$ AU. This does not set a rigorous upper limit on $a_{\rm ABC}$, but we believe it is reasonably conservative.

\begin{figure}
\centering
\includegraphics[width=1.0\linewidth]{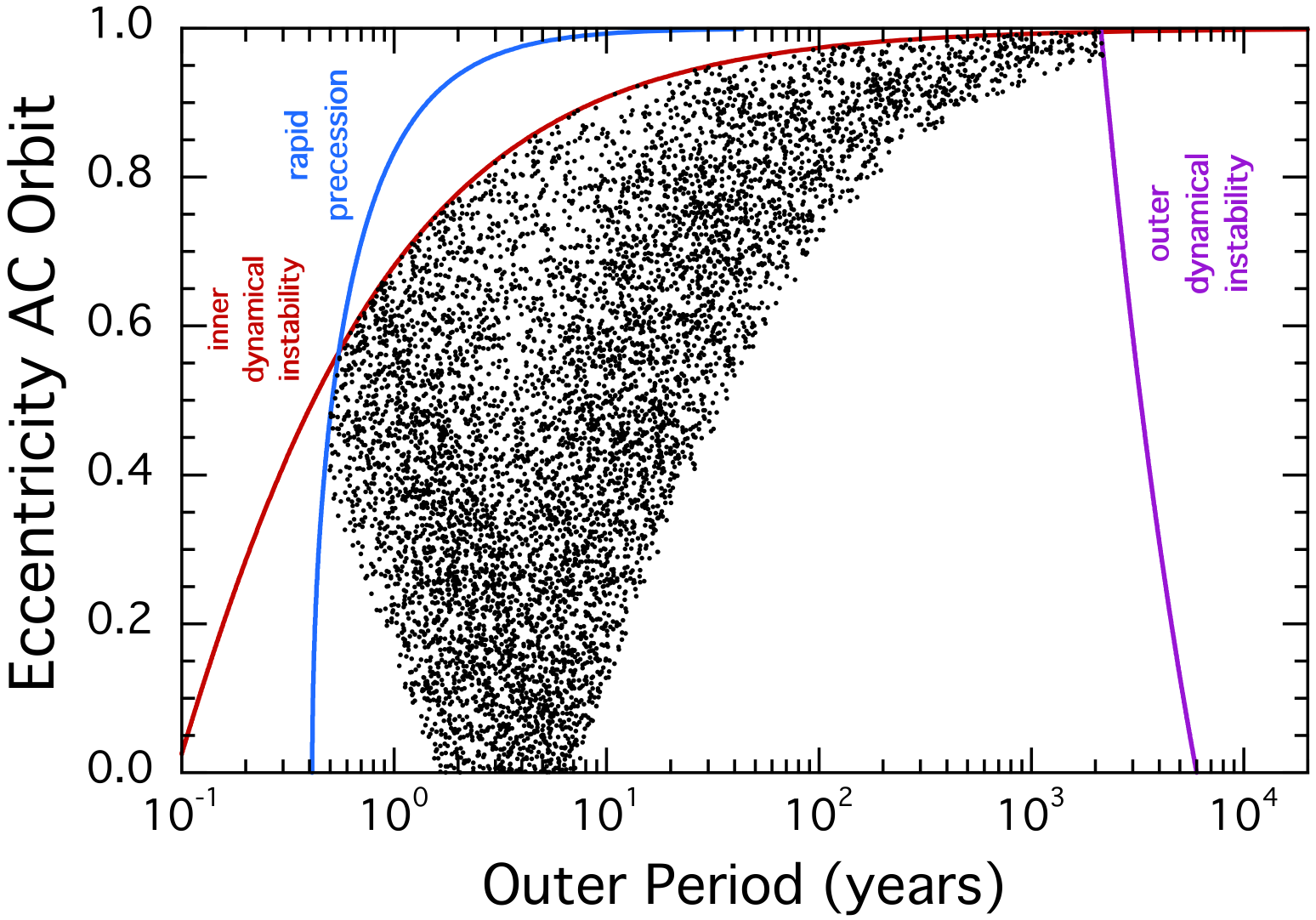}
\caption{Allowed orbits in the $P_{\rm AC}-e_{\rm AC}$ plane.  The analytic constraints are shown by the colored curves and are discussed in the text.  Additionally, the orbits are required to produce a motion of star A with respect to an essentially motionless star B, yielding $\sim$8 km s$^{-1}$ in the radial direction and a relative PM of $\sim$5 mas yr$^{-1}$.}
\label{fig:ecc_vs_P}
\end{figure}

\begin{figure}
\centering
\includegraphics[width=1.0\linewidth]{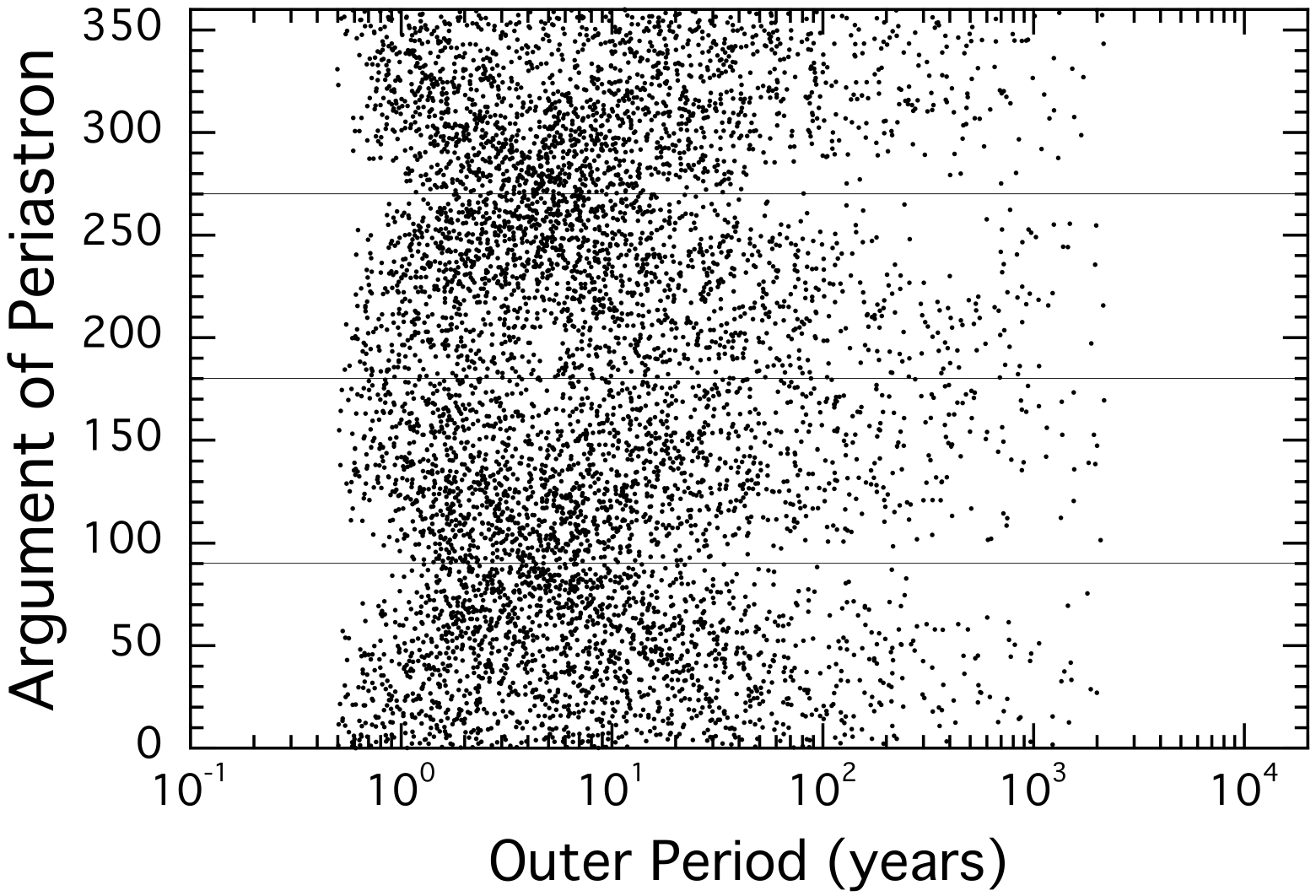}
\caption{Allowed orbits in the $P_{\rm AC}-\omega_{\rm AC}$ plane.}
\label{fig:omega_vs_P}
\end{figure}

Our approach was to randomly select orbital parameters for the AC orbit uniformly in $\log P_{\rm AC}$, from 1 to $10^7$ days; uniformly in eccentricity from 0 to unity; uniformly in argument of periastron, $\omega$ from 0 to $2 \pi$; and orbital inclination according to $\cos^{-1}(\mathcal{R})$, where $\mathcal{R}$ is a draw from uniform random number generator.  We followed each of $10^5$ trial orbits around a complete orbit.  Constraints (i), (ii), and (iii) listed above were immediately applied to each trial orbit, and if any of those tests failed, that particular orbit was rejected.  For constraints (iv) and (v) listed above, those conditions were checked everywhere around an entire orbit.  If they were never satisfied, the orbit was also rejected.

The results of acceptable AC orbits are shown in Figs.~\ref{fig:ecc_vs_P} and \ref{fig:omega_vs_P}.  In Figure~\ref{fig:ecc_vs_P} we show the orbits which satisfy all five constraints in the $P_{\rm AC} - e_{\rm AC}$ plane.  Since constraints (i)-(iii) are analytic functions of $P_{\rm AC}$ and $e_{\rm AC}$, and a reasonable assumption about $P_{\rm ABC}$ (see above), we show those as solid limiting curves.  The remainder of the constraints come from the RV and PM differences between stars A and B.  It seems clear that inner orbital periods $P_{\rm AC}$ ranging from about a year to 1000 years are acceptable. All eccentricities seem possible. However, for nominal eccentricities in triple systems of $\sim$$0.2-0.7$ (see Figure~9 of \citealt{Borkovits2016}) we more likely can expect to find $P_{\rm AC}$ in the range of $1-50$ yr. In Figure~\ref{fig:omega_vs_P} we show $\omega_{\rm AC}$ vs $P_{\rm AC}$.  These two parameters are slightly correlated, with more allowed orbits having $\omega_{\rm AC} \simeq 90^\circ ~{\rm and}~270^\circ$ and covering a more restricted range of $P_{\rm AC}$.  The inclination angles $i_{\rm AC}$ (not shown in the plots) are uniformly distributed between a minimum of $68^\circ$ and $90^\circ$.

\section{Comparison with Other Low-Mass Eccentric Binaries}

Similar to the case of the vast majority of close main-sequence binaries, it is clear that since the radii of the low-mass protostars were much larger than the present-day separation of its stars, the eccentric binary Cab cannot have been formed in its present orbital configuration \citep[see, e.g.][]{Kiseleva1998,Fabrycky2007}. Perhaps the most interesting question about this close pair is how it has managed to retain such a high eccentricity despite the requisite orbital shrinkage during its past history. Some possibilities are as follows: (i) the initial eccentricity of the originally wider orbit was much higher, and while there is ongoing tidal friction causing orbital shrinkage and circularization, the system is still sufficiently young that there has not been enough time to circularize the orbit; (ii) the shrinkage of the initial orbit, which was wide enough to accommodate the protostars, was the result of some other mechanism(s) beside tidal dissipation such as, e.g., (a) escape of an additional stellar component or, (b) accretion-driven migration \citep[see, e.g.][and further references therein]{TokovininMoe2020}. Or, another possibility is, (iii) that the observed current high eccentricity is a consequence of ongoing dynamical interactions with the more distant, third and fourth stellar components of the quadruple system.

\begin{figure}
\includegraphics[width=1.0\columnwidth]{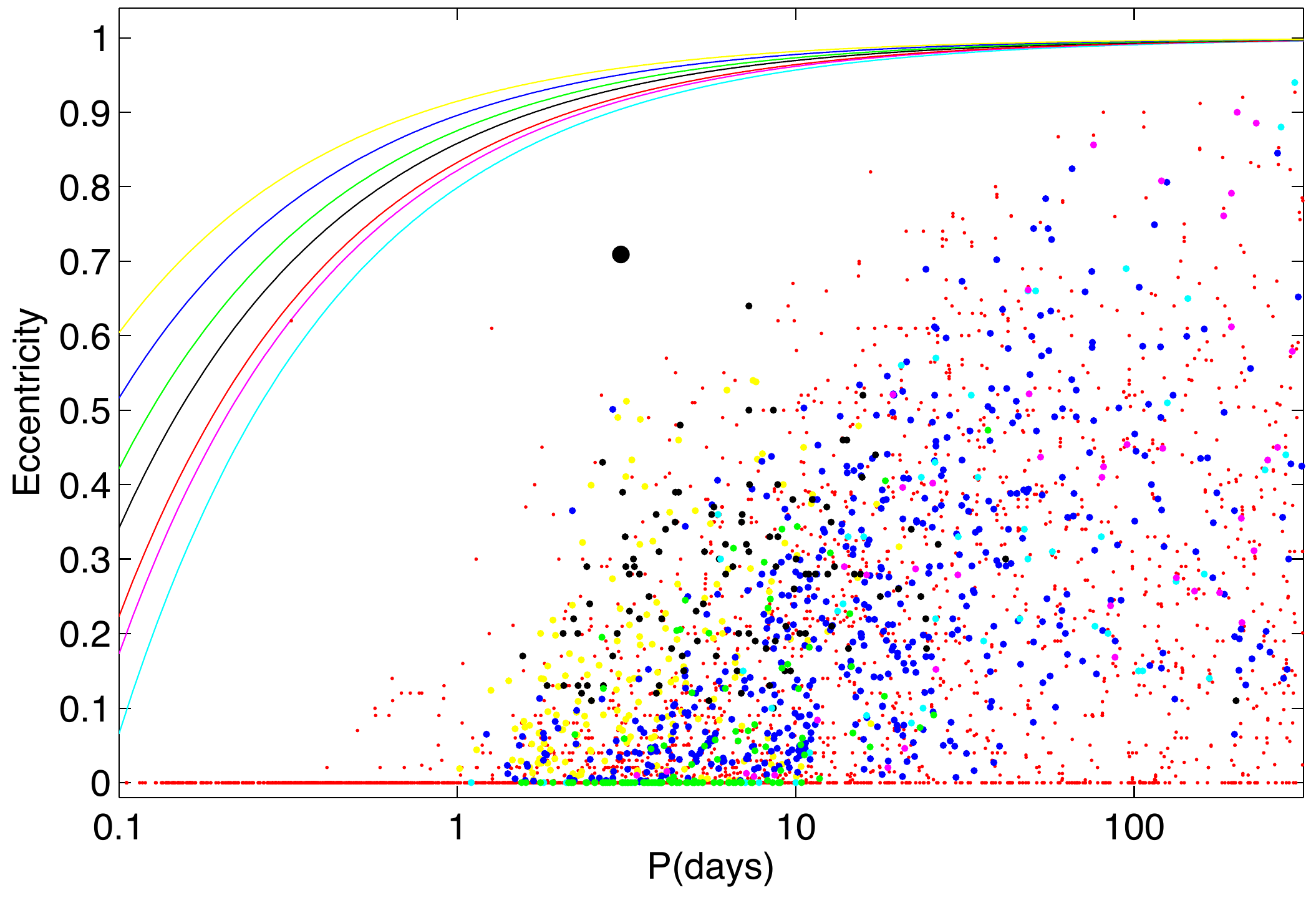}
\includegraphics[width=1.0\columnwidth]{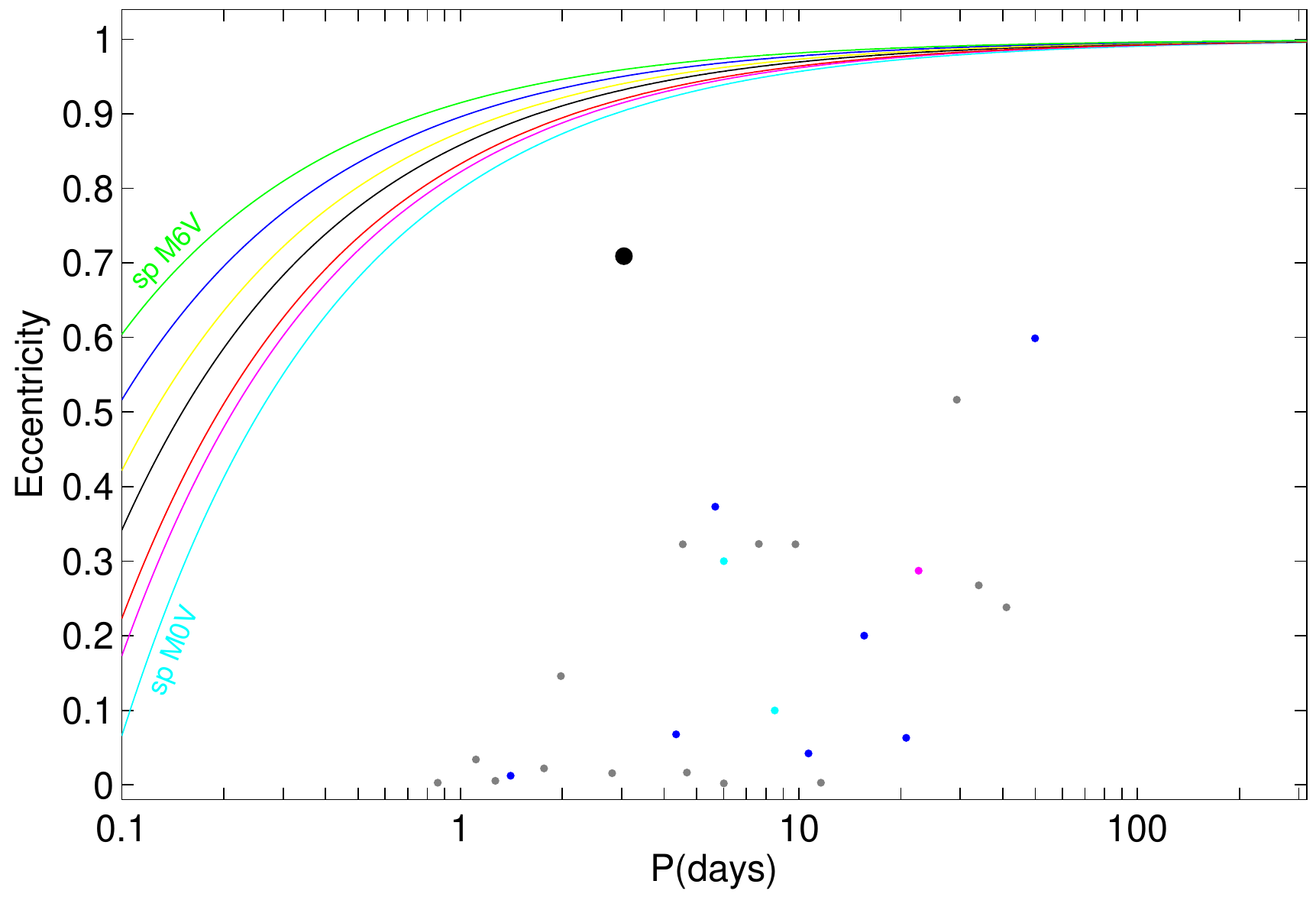}
\caption{Empirically determined orbital periods and eccentricities of main-sequence binaries from \citet{zasche21}. {\em Top panel:} all stars; {\em Bottom panel:} only low mass stars of the M spectral type (see text for references). The heavy black circle represents the inner eccentric binary in TIC 121088960.  The colors of the other dots are from different studies and are listed in the text. The smooth colored curves are upper bounds that are set by when near collisions between the stellar components of the binary would occur at periastron.
}
\label{fig:P_e}
\end{figure}

We constructed a so-called ``$P-e$'' diagram showing the known cases of eccentric binaries in Figure \ref{fig:P_e} (upper panel). We plotted there the same datasets as in a previous paper by \citet{zasche21}: small red dots are from the SB9 catalogue \citep{pourbaix04}; yellow dots show eclipsing binaries from the catalogue of eccentric binaries by \cite{kim18}; blue dots show Kepler binaries by \citet{kjurkchieva17}; black dots show those from ASAS published by \cite{shivvers14}; cyan points are from \citet{halbwachs03}; green points are from \citet{triaud17}; and magenta points are from \citep{latham02}.  All of these were studied as spectroscopic and/or eclipsing binaries.  However, some of the most extreme points from the SB9 catalogue have very uncertain orbits and should not be considered as real eccentricities. Moreover, most of the data shown here represent much more massive stars, for which the circularization process is different due to their internal structure (as recently proven on real data, see e.g. \citealt{vaneylen16}).

For these reasons, we have tried to compare the comparable, i.e., plot our unique low-mass system with other stars having derived eccentricities and both components of the M spectral type. Specifically, (i) the set of {\em Kepler} eclipsing binaries yielded 7 systems with GAIA photometric index $(B_p-R_p) > 1.8$ mag; (ii) the spectroscopic survey by \citet{latham02} contains only one system with effective temperature below $4000$ K; and (iii) the survey \citet{halbwachs03} provided two binaries with masses below 0.6~$M_\odot$. These are plotted in the bottom panel of Fig.~\ref{fig:P_e}. These data are then complemented with other M+M binaries taken from various dedicated studies (\citealt{Stassun2006}; \citealt{Morales2009b}; \citealt{Carter2011}; \citealt{Irwin2011}; \citealt{Gomez2012}; \citealt{Kraus2015}; \citealt{David2016}; \citealt{Gillen2017}; \citealt{Kraus2017}; \citealt{Lubin2017}; \citealt{Irwin2018}; \citealt{Murphy2020}; \citealt{Acton2020}). The data for these low-mass binaries are plotted together in the bottom panel of Fig.~\ref{fig:P_e} with curves representing the close pericenter approaches (i.e., $1.5 \times R_\star = a \cdot (1-e)$) when they likely collide with each other. The periastron separations are calculated for different spectral types (from M0V to M6V) according to their typical radii and masses according to \cite{pecaut13}, assuming both components are similar to each other (same masses and radii). One can also ask whether some proximity effect near periastron passage would also be visible on the light curve of the binary. However, we have calculated that such an effect is so small that it cannot be detected with the current precision of the available {\em TESS} data.

As one can see from the $P-e$ plot, our system is located in the very upper part of the diagram. Hence, any statement about a fixed circularization period of about 20 days \citep{latham02}, or about 10 days \citep{meibom05}, or 4 days \citep{halbwachs03} should not be taken literally. Probably the most suitable for our system is the discussion about the circularization period by  \citet{triaud17}.  They pointed out that these values derived from clusters are not in agreement with their findings based on analysing 118 F/G/K + M-type binaries during their ``EBLM Project''. \citet{halbwachs03} found that the twins (i.e., those systems having $q > 0.8$) tend to have lower eccentricities compared to non-twins. However, our system which has a mass ratio close to 0.8 for the binary components ($q = 0.73$) indeed has a very significant eccentricity, so any such statement should be taken with caution.

Most recently \citet{Justesen2021} have investigated the stellar temperature and separation dependence of tidal circularization on the large sample of eclipsing binaries observed by the {\em TESS} satellite in the southern hemisphere. Their sample contains 58 binaries similar to our innermost pair Cab in that all the stars are cooler than 4\,500 K and, therefore, similar to our stars, they are highly convective. Therefore, we assume that the results found for those systems should be relevant in the context of the innermost pair Cab. \citet{Justesen2021} computed the critical scaled distance (i.e., $a/R$) for circularization of these systems at the {\em end} of their pre-main-sequence phase using the quantitative predictions of the revised equilibrium tide theory of \citet[][Table 2]{zahn89}. Then, they compared these computed values with the ones they inferred from the corresponding sample of {\em TESS} EBs.  For convective stars, they found good agreement.

More specifically, \citet{Justesen2021} found that for the lowest-mass stars ($T_\mathrm{eff}<4\,500$\,K) the observered critical scaled distance is $(a/R)_\mathrm{crit}^\mathrm{obs}=29.8_{-4}^{+42}$, in agreement with the theoretical value of $(a/R)_\mathrm{crit}^\mathrm{theo}=28-32$. Comparing these results with our eccentric EB, Cab, by taking the binary parameters from Table~\ref{tbl:MRTL} above, one can see that $a/R_1\approx30$ for the primary component.  Turning to the orbital periods instead of scaled distances, \citet{Justesen2021} found that amongst the coolest binaries in their sample, the circularization period (i.e., the orbital period below which binaries are expected to be circularized) was found to be $P_\mathrm{crit}^\mathrm{obs}=5.57_{-0.66}^{+0.20}$\,d. Therefore, we can conclude that according to both the empirical statistical investigations of \citet{Justesen2021} and the revised equilibrium tide model of \citet{zahn89}, in the absence of any other eccentricity-exciting mechanism, tidal effects might have circularized the innermost orbit during the pre-main-sequence phase of their stellar evolution. Therefore, we can reasonably assume that the currently observed high eccentricity of Cab is not primordial, but rather a consequence of some later dynamical effect. In what follows we investigate this possibility.

\section{The Effect of von Zeipel-Lidov-Kozai Cycles}
\label{sec:KL}
In a hierarchical triple (or, multiple) stellar system the orbital motions of the components no longer remain purely Keplerian. Therefore, the orbital elements, including the eccentricity of the inner close binary, will continuously vary in time due to the perturbations of the third (and even further) bodies. In a hierarchical stellar configuration the perturbations to each orbital element of the inner binary have three characteristic timescales, proportional to the inner binary period ($P_\mathrm{in}$), the outer orbital period ($P_\mathrm{out}$), and to their ratio in the form of $P_\mathrm{out}^2/P_\mathrm{in}$. While the former two classes of short-period perturbations generally result in small-amplitude variations in each orbital element \citep[see, e.g.][]{Borkovits2015}, the long period ones that are usually referred as either `secular' or `apse-node' timescale perturbations may result in substantial variations in the configuration of the given system.  These secular effects, pioneered by \citet{vonZeipel1910}, \citet{Lidov1962}, and \citet{Kozai1962}, have recently been referred to as the `von Zeipel-Lidov-Kozai effects' (hereafter ZLK) or `ZLK oscillations', and are the subject of many studies over the last two and half decades. Detailed reviews of the ZLK phenomena can be found in \citet{Naoz2016} and \citet{Ito2019}.

In its general formulation, a satisfactory analytical description of ZLK oscillations requires consideration of higher order (at least octupole) terms of the secular perturbation function. The octupole terms become important only for small outer to inner period ratios and highly eccentric outer orbits, i.e., when one cannot assume that the vast majority of the orbital angular momentum of the triple system (or, in the present situation, of the inner triple subsystem -- AC) is stored in the outer orbit. For a visualisation of the parameter regime where the octupole terms are important, for a given set of triple star masses, see, e.g., Figure~4 of \citet{Toonen2020}. Strictly speaking, in the absence of any information about the period ratio and the outer eccentricity of the triple subsystem (AC), we cannot decide whether the octupole effects have any significance in the current system. We are, however, convinced that for our exploratory quantitative analysis, the use of the much simpler quadrupole approximation is perfectly illustrative. The greatest advantage of this latter quadrupole approximation is that it has only one degree of freedom and thus is integrable. Furthermore, it has analytic solutions which, regarding the hierarchical triple-star (AC) dynamics, have been extensively discussed by \citet{Harrington1968} and \citet{Soderhjelm1982}. According to these discussions, there are two domains of the mutual inclination angle of a hierarchical triple system limited by the value of $\cos^2i_\mathrm{mut}\approx\frac{3}{5}$; in these two domains there is a substantial difference in the nature of the cyclic eccentricity variations of the inner binary. (Note, an exact equivalence occurs only in the case of the asymptotic solution, i.e., when the orbital angular momentum is stored exclusively in a circular outer orbit.)

For systems with low mutual inclinations (i.e. $i_\mathrm{mut}\lesssim39\fdg23$ or, $i_\mathrm{mut}\gtrsim140\fdg77$) there may occur only small-amplitude secular variations in the inner eccentricity (and the mutual inclination, as well). Furthermore, in the case of an initially circular inner orbit, it does remain circular at every prior and later time (of course, only as far as the approximation used remains valid). From our perspective, the high mutual inclination regime is much more interesting.  In this case, depending on the initial conditions, (i.e., the value of the inner binary eccentricity, mutual inclination and dynamical argument of pericenter at a given instant), the inner eccentricity may vary between zero and nearly unity, while the apsidal line may exhibit either circularization or libration.\footnote{Note, however, one can find specific solutions even for nearly perpendicular configurations, as well, where the inner eccentricity remains (essentially) constant and the major axis of the orbit freezes into a specific direction.} The most interesting case for us occurs when a nearly circular orbit periodically becomes an extremely eccentric one. In this regard one can show that, as far as $e_{\rm in}^\mathrm{min}\rightarrow0$, then, assuming that the asymptotic approximation is valid,
\begin{equation}
e_{\rm in}^\mathrm{max} \rightarrow \sqrt{1-\frac{5}{3}\left(1-e_\mathrm{in}^\mathrm{min}\right)^2\cos^2i_{\rm mut}^\mathrm{min}}~~,
\end{equation}
while the corresponding mutual inclination at the time when the inner orbit achieves its maximum eccentricity becomes
\begin{equation}
\cos i_\mathrm{mut}^\mathrm{max}=\pm\sqrt{\frac{3}{5}}.
\end{equation}
(Note, $\cos i_\mathrm{mut}^\mathrm{min,max}$ refer to the values of $\cos i_\mathrm{mut}$ at those instants when the inner eccentricity takes its minimum and maximum values, i.e., $e_\mathrm{in}^\mathrm{min,max}$, respectively, and is not to be confused with the minimum and maximum values of $\cos i_\mathrm{mut}$, itself.)
It can also be shown that, when the inner orbit has its maximum eccentricity, the dynamical argument of pericenter has a value of $g_\mathrm{in}=\pm90\degr$ (i.e., the semi-major axis is perpendicular to the intersection of the inner and outer orbital planes). Therefore, in order to generate such an orbital configuration for our triple system AC (i.e., TIC~121088960) where the inner eccentricity oscillates between (almost) zero and the currently observed high value of $e_\mathrm{in}\approx0.7$, we have to simply set the present mutual inclination between the inner and outer (triple) orbits to about $i_\mathrm{mut}\approx40\degr$ (or, its retrograde counterpart), and the dynamical argument of periastron to $g_\mathrm{in}\approx\pm90\degr$. Naturally, there is no reason to assume that the currently observed value of $e_\mathrm{in}\approx0.7$ belongs to the maximum eccentricity phase of an ongoing ZKL cycle.  However, as our purpose is simply to illustrate that the ZKL effect by itself may produce the observed high eccentricity of our system, this assumption is perfectly satisfactory for our goals, and makes it easier to set the initial parameters for the illustrative numerical integrations shown below.

We can also investigate in an approximate, but fully analytic, approach the characteristic timescale of these large amplitude eccentricity cycles.  In this regard, \citet{Soderhjelm1982} gives an analytical formula which depends on the mass and period ratios, the eccentricity of the outer orbit, and also on the amplitude of the eccentricity cycles as follows:
\begin{equation}
\tau_\mathrm{quad}\approx\frac{4}{3\sqrt6}\frac{m_\mathrm{A}+m_\mathrm{Ca}+m_\mathrm{Cb}}{m_\mathrm{A}}\frac{P_\mathrm{out}^2}{P_\mathrm{in}}\left(1-e_\mathrm{out}^2\right)^{3/2}\times f(\Delta e_\mathrm{in}),
\label{eq:ZKL_timescale}
\end{equation}
where the functional dependence on $e_{\rm in}$ is simply denoted as $f(\Delta e_\mathrm{in})$.\footnote{The calculation of $f(\Delta e_\mathrm{in})$ is rather lengthy, though straightforward, and therefore we do not repeat it here. The formulation can be found in Eqs.~(29)--(32) of \citet{Soderhjelm1982}.} This expression, evaluated for the masses given in Table~\ref{tbl:MRTL}, gives
\begin{equation}
\tau_\mathrm{AC}\approx864.2\times\left(\frac{P_\mathrm{out}}{1000\,\mathrm{d}}\right)^2\left(1-e_\mathrm{out}^2\right)^{3/2}\times f(\Delta e_\mathrm{in})~\mathrm{[yr]}.
\end{equation}

\begin{figure}
\centering
\includegraphics[width=1.0\linewidth]{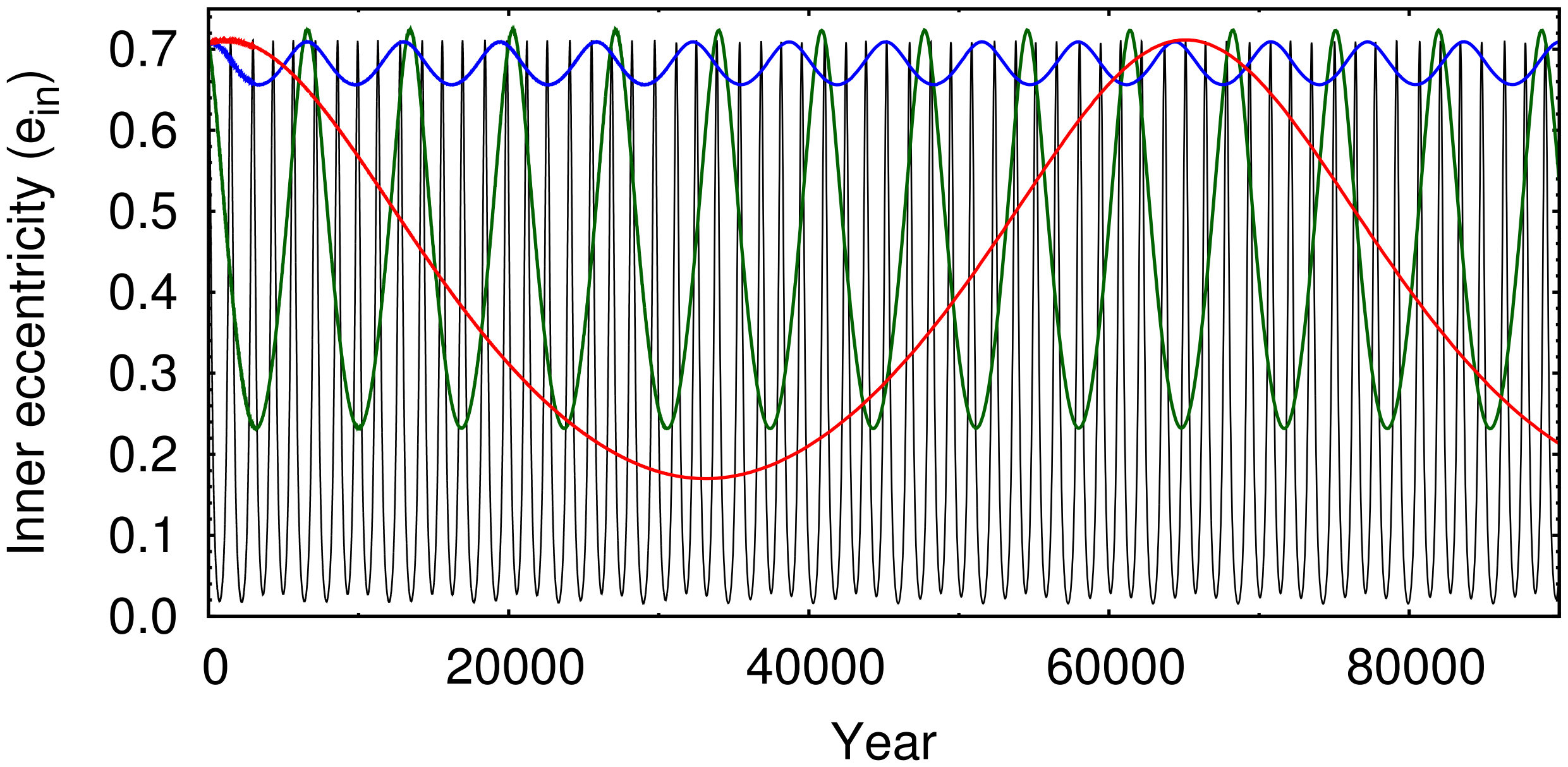}
\caption{Examples of ZKL forced large-amplitude eccentricity oscillations of the inner binary C.  Black, green, blue and red curves belong to the runs of which some characteristic parameters are tabulated in columns $\#1-4$ of Table~\ref{tbl:numint}. (See text for a detailed discussion.)}
\label{fig:ZKL_ecc}
\end{figure}

To illustrate the validity of the approximations we have used, we carried out some numerical integrations. We applied the numerical integrator described in \citet{Borkovits2004}. Besides the gravitational point-mass three-body interactions we considered the tidal effects that arise between the two stars of the close eccentric binary (C). For these runs, the masses were taken from Table~\ref{tbl:MRTL}, while the current orbital elements of the eccentric EB were taken from our lightcurve solution (Table~\ref{tab:fitted_parameters}). The period ($P_\mathrm{out}$) and eccentricity ($e_\mathrm{out}$) of the outer  orbit of triple system AC were chosen arbitrarily, but in such a manner so as to be in accordance with the parameters of the allowed orbits (see Figs.~\ref{fig:ecc_vs_P} and \ref{fig:omega_vs_P}). Moreover, we set the plane-of-the-sky-related outer inclination, $i_\mathrm{out}$, and the difference of the plane-of-the-sky-related longitude of the nodes ($\Delta\Omega=\Omega_\mathrm{out}-\Omega_\mathrm{in}$) in such a manner as to provide the necessary values for the dynamical frame of reference related quantities (i.e., $g_\mathrm{in}\approx\pm90\degr$ and $i_\mathrm{mut}\approx40\degr$ or, $i_\mathrm{mut}\approx140\degr$).  Finally, the strength of the tidal interaction was controlled with the usual apsidal motion parameter $k_2$. We tabulate the initial parameters and some characteristic quantities of the `measured' ZKL eccentricity cycles for a number of illustrative runs in Table~\ref{tbl:numint}, while the variations of $e_\mathrm{in}$ from its present value are plotted in Figure~\ref{fig:ZKL_ecc}.

As one can see, for a relatively short orbital period of the AC subsystem ($P_\mathrm{AC}=1000$\,d), one can readily find an orbital configuration that satisfies the constraints discussed in Sect.~\ref{sec:tripleorbit}, and which also results in a rapid, practically continuous, variation of the inner eccentricity between (almost) zero, and the currently observed high value of $e_\mathrm{in}\sim0.7$ on a `short' timescale of about a millennium (first column in Table~\ref{tbl:numint}, and black curve in Figure~\ref{fig:ZKL_ecc}).

Choosing an order of magnitude longer outer period ($P_\mathrm{AC}=10\,000$\,d), however, one can notice another interesting aspect of the ZKL cycles. In the case of run $\#3$ (see Table~\ref{tbl:numint}), despite the fact that the input parameters were set such that one might again expect large amplitude ZKL cycles, the numerical integration instead resulted in only a small amplitude cyclic variation in $e_\mathrm{in}$ (and, accordingly, in $i_\mathrm{mut}$, as well). The reason is that, for this configuration, the rate of the tidal-oblateness-generated apsidal motion of binary Ca-Cb has exceeded that of the third-body forced apsidal motion, and this effect has killed the ZKL cycles.\footnote{According to our knowledge, it was \citet{Soderhjelm1984} who reported for the first time that sufficiently strong tidal oblateness can eliminate the ZKL effect. Later, the question was elaborated in more detail in the seminal works of \citet{Eggleton1998,Kiseleva1998} which have led to the theory of KCTF.} In order to illustrate this, in the case of run $\#4$ we used similar input parameters as in run $\#3$, but the apsidal motion constant $k_2$ was set to $k_2^\mathrm{Ca,Cb}=0.0001$, i.e., for all practical purposes we switched off the tidal effects. In that case, the ZKL cycles return, but naturally on a much longer timescale due to the longer $P_\mathrm{AC}$.

\begin{table}
\centering
\caption{Initial orbital elements for the numerical integrations that serve as examples of large amplitude ZKL cycles. The first set of the tabulated parameters (from $P_\mathrm{out}$ to $k_2^\mathrm{(Ca,Cb)}$) are the adopted current values for the triple AC. The next three parameters are dynamical orbital elements ($i_\mathrm{mut}$, $g_\mathrm{in}$ and $g_{\rm out}$) and are calculated for the epoch time from the usual orbital elements. $P_e$ is the actual period of the eccentricity oscillations or, more strictly speaking, the time elapsed between the first and second eccentricity maxima. The final two parameters are $e_\mathrm{in}^{\rm min,max}$ which are the minimum and maximum values of the eccentricity of the inner binary during the first integrated ZKL cycle.}
\begin{tabular}{lcccc}
\hline
&  $\#1$ & $\#2$ & $\#3$ & $\#4$ \\
\hline
$P_\mathrm{out}$ [d] & $1000$ & $3652.5$ & $10000$ & $10000$  \\
$e_\mathrm{out}$     & $0.35$ & $0.41$ & $0.50$  & $0.50$   \\
$\omega_\mathrm{out}$ [$\degr$] & $90$ & $104$ & $270$ & $270$ \\
$i_\mathrm{out}$ [$\degr$] & $84$ & $98$ & $84$ & $84$ \\
$\Delta\Omega$ [$\degr$] & $40$ & $45$ & $140$ & $140$ \\
$k_2^\mathrm{(Ca,Cb)}$ & $0.0200$ & $0.0200$ & $0.0200$ & $0.0001$ \\
\hline
$i_\mathrm{mut}$ [$\degr$] & $40$ & $46$ & $139$  & $139$  \\
$g_\mathrm{in}$ [$\degr$]& $267$ & $287$ & $264$  & $264$ \\
$g_\mathrm{out}$ [$\degr$]&$355$ & $24$ & $189$  & $189$ \\
$P_e$ [yr] & $1470$ & $6830$ & $6440$  & $63900$ \\
$e_\mathrm{in}^\mathrm{min}$ & $0.019$ & $0.232$ & $0.656$ & $0.170$ \\
$e_\mathrm{in}^\mathrm{max}$ & $0.712$ & $0.724$ & $0.709$ & $0.712$ \\
\hline
\label{tbl:numint}
\end{tabular}
\end{table}

Up to this point we have concentrated only on the triple subsystem AC. However, the presence of the more distant component B makes things more complicated. For example one can imagine a situation where the inner triple AC would have been originally a nearly flat and circular system, while ACB are in an orbit that is highly inclined to the plane of AC. Let's consider periods $P_\mathrm{AC}=10$\,yr and $P_\mathrm{ABC}=6000$\,yr. In this case, according to Eq.~(\ref{eq:ZKL_timescale}) one may expect high amplitude ZKL oscillations on a timescale of some Myrs. These may lead not only to a high eccentricity of the middle orbit ($e_\mathrm{AC})$ but, what is more crucial, to a substantial change in the inclination of the middle orbit ($i_\mathrm{AC}$) which, in turn, might switch on the ZKL cycles in the inner triple subsystem.

In conclusion, as one can see, the dynamical history of such a 2+1+1 quadruple star system may be extremely rich and interesting.  The currently observed high eccentricity of the innermost close pair (binary C) can definitely be the consequence of an ongoing ZKL process.

\begin{figure}
\centering
\includegraphics[width=1.0\linewidth]{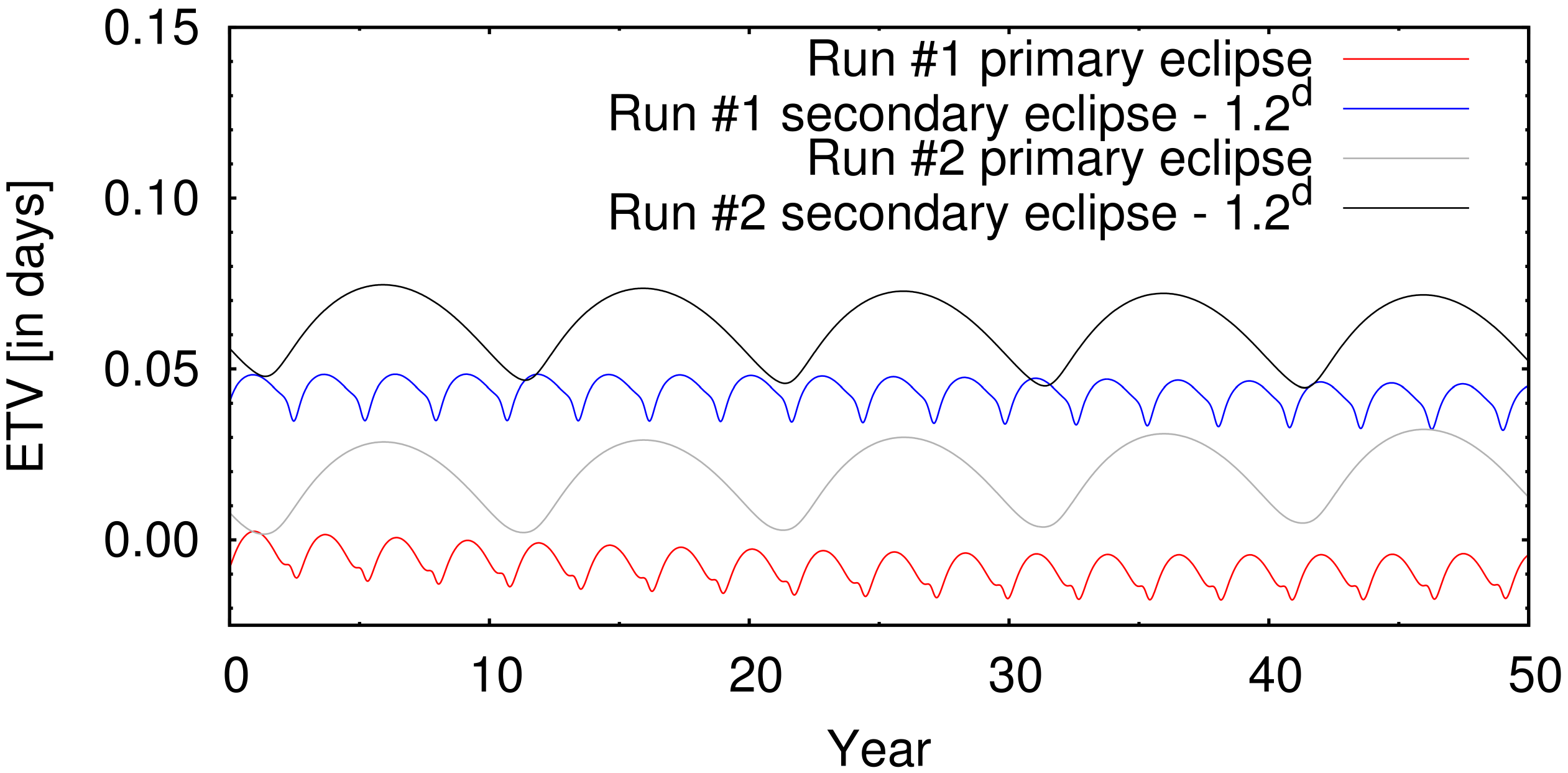}
\caption{Simulated ETV curves for runs $\#1$, $\#2$ for the next 50 years. As expected, the ETVs for both a $P_\mathrm{AC}=1000$\,d and a $P_\mathrm{AC}=10$ yr period should be detectable within years or, even months. These show the non-linear behavior in the ETVs on the timescale of the orbit of the triple AC.}
\label{fig:ZKL_ETV}
\end{figure}

The question naturally arises as to whether there are any observational consequences of such a process and, if so, on what timescales they will be detectable. In order to investigate this question we plot the numerically generated Eclipse Timing Variations (ETV) curve for runs $\#1$ and $\#2$ in Figure~\ref{fig:ZKL_ETV}. As one can see, in the case of a relatively short-period outer orbit (i.e., 3-10 years), ETVs should be detectable within years or even  months. Moreover, due to the fact that the outer body (A) is expected to be highly inclined and, furthermore, during an ongoing high-amplitude ZKL cycle the inclination of the innermost binary (C) varies very quickly, we can also expect substantial eclipse depth variations within a few years. Therefore, the continuous monitoring of the eclipses of this intriguing system is highly encouraged.

\section{Conclusions}

We report the discovery of a hierarchical 2+1+1 quadruple stellar system discovered with {\em TESS}, containing TIC 121088959 and TIC 121088960 at an angular separation of $\sim$3.9$''$ and a highly-eccentric EB as an unresolved part of TIC 121088960. We analyzed both the S4 long-cadence and S31 short-cadence {\em TESS} eclipse photometry to measure the properties of the component stars of the EB. Our analysis shows that the EB has a highly-eccentric ($e$ = 0.709) short-period ($P = 3.04358$-day) orbit, which occupies an extreme region of eccentricity-period phase space for eccentric binaries.  We made use of the {\em TESS} light centroid in the difference image (in-eclipse vs.~out-of-eclipse regions) to show that the most likely host of the EB is TIC 121088960, while TIC 121088959 is ruled against at the 3-$\sigma$ level.

\indent To learn more about the system, we performed spectroscopic follow-up observations using the iSHELL at NASA's Infrared Facility and the Coud{\'e} spectrograph at the McDonald 2.7 m telescope and measured the RVs of both TIC 121088959 and TIC 121088960. Our RV measurements show no changes in RV over an 11-day baseline for both stars and the difference in RV between the two stars is 8 $\pm$ 0.3 \kms.

Gaia measured PMs and parallaxes for both TIC 121088959 and TIC 121088960, and these are listed in Table \ref{tab:mags}. The difference in the PMs is $5 \pm 0.1$ mas yr$^{-1}$ which corresponds to 2 km s$^{-1}$ at the distance of the two stars, which is close to 83 pc. The Gaia measurements provide a strong indication that the two objects, TIC 121088959 and TIC 121088960, are gravitationally bound.

The shallow depths of the eclipses ($\lesssim 3\%$) and constant RVs both suggest that the light from the eccentric eclipsing binary is being substantially diluted by other light in one of the two close stellar images.  The Gaia astrometric excess noise and RUWE values for TIC 121088960 indicate that its astrometric solutions are being affected by more than one star.  This, plus the fact that light centroids of the eclipses (see above) point toward TIC 121088960, indicate that this object is the host of the eccentric EB plus another star.

To further analyze the system, we performed SED fits using Gaia and 2MASS photometric data.  We thereby estimated the stellar masses of the four stars, all of which are in the M dwarf regime (see Sect. \ref{sec:PhotConstraints}). Specifically, the stars in the eccentric EB are of approximately M5V spectral type, while the two directly visible resolved stars are of M2.5V--M3V spectral type.  This makes the unresolved EB the most eccentric and short period M dwarf EB known-to-date (see Figure \ref{fig:P_e}).

To estimate the orbital elements of the inner triple system, we performed numerical simulations covering a grid in period, eccentricity, argument of periastron, and inclination (see Sec. \ref{sec:tripleorbit}). Our simulations led us to conclude that the orbital period of the inner triple is between 1 to 1000 years at the extreme, and much more likely to lie in the range 1--50 years.  All that we can say about the outer quadruple orbit is that the period is $\sim10^4$ years.

In Sect.\ref{sec:KL}, we investigated the likelihood that von Zeipel-Lidov-Kozai cycles would be able to produce long intervals where the eccentricity of the inner EB (C) is kept high.  Based on an analytic expression, we showed that ZLK cycles would have characteristic timescales in this system of $\approx 860 (P_{\rm out}/1000 d)^2$ years, and indeed we might be viewing this system in one of its cyclicly recurring high-eccentricity states. We went on to numerically integrate a few different examples of ZLK cycles in this system and confirm the expectations based on the analytic expressions.  We also simulated ETV curves for the next few decades and showed that for a plausible range of periods for the AC triple (i.e., a few years), we can expect measurable non-linear perturbations to the ETV curves on the timescale of the triple period.  Finally, we encouraged monitoring of the ETVs and eclipse depths in this system since these changes may well be detectable within a few years (or even shorter).

\section*{Acknowledgements}
The authors are grateful to an anonymous referee for a very careful review and many excellent suggestions for substantial improvement of our presentation.
We wish thank Andrei Tokovinin for his extremely helpful comments on the manuscript.
We also wish to acknowledge Jon Jenkins (NASA Ames Research) and Joseph Twicken (SETI Institute) for important input concerning the implementation of {\em TESS} light centroiding.
Our visual survey group (VSG) acknowledges Allan R. Schmitt for making his lightcurve examining software LcTools freely available.
This paper includes data collected by the TESS mission. Funding for the TESS mission is provided by the NASA Science Mission directorate. Some of the data presented in this paper were obtained from the Mikulski Archive for Space Telescopes (MAST). STScI is operated by the Association of Universities for Research in Astronomy, Inc., under NASA contract NAS5-26555. Support for MAST for non-HST data is provided by the NASA Office of Space Science via grant NNX09AF08G and by other grants and contracts.

This publication makes use of data products from the Two Micron All Sky Survey, which is a joint project of the University of Massachusetts and the Infrared Processing and Analysis Center/California Institute of Technology, funded by the National Aeronautics and Space Administration and the National Science Foundation.

\section*{Data Availability}

The {\it TESS} data presented in this work were obtained from the Mikulski Archive for Space Telescopes(MAST) portal (https:
//mast.stsci.edu/portal/Mashup/Clients/Mast/Portal.html). The photometric data presented in Table 1 were obtained from sources in public domain which are given in the caption. Facilities that are used for the work are {\it TESS}, Gaia, IRTF (iSHELL), and McDonald observatory (Robert G. Tull Coud{\'e} spectrograph).



\bibliographystyle{mnras}
\bibliography{main}

\end{document}